# Geometry and Astronomy: Pre-Einstein Speculations of Non-Euclidean Space

HELGE KRAGH
Centre for Science Studies, Department of Physics and Astronomy
Aarhus University, 8000 Aarhus, Denmark
E-mail: helge.kragh@ivs.au.dk

**Abstract** The recognition that physical space (or space-time) is curved is a product of the general theory of relativity, such as dramatically shown by the 1919 solar eclipse measurements. However, the mathematical possibility of non-Euclidean geometries was recognized by Gauss more than a century earlier, and during the nineteenth century mathematicians developed the pioneering ideas of Gauss, Lobachevsky, Bolyai and Riemann into an elaborate branch of generalized geometry. Did the unimaginative physicists and astronomers ignore the new geometries? Were they considered to be of mathematical interest only until Einstein entered the scene? This paper examines in detail the attempts in the period from about 1830 to 1910 to establish links between non-Euclidean geometry and the physical and astronomical sciences, including attempts to find observational evidence for curved space. Although there were but few contributors to "non-Euclidean astronomy," there were more than usually supposed. The paper looks in particular on a work of 1872 in which the Leipzig physicist K. F. Zöllner argued that the universe is closed in accordance with Riemann's geometry.

1. *Introduction*
2. *Non-Euclidean geometries of space*
3. *Astrophysicist, controversialist, spiritualist*
4. *Zöllner's closed universe*
5. *Responses and criticism*
6. *Some later ideas, 1870-1900*
7. *The pre-relativistic cosmos*
8. *Towards Einstein*
9. *Conclusion*

## 1. Introduction

Ever since the beginning of relativistic cosmology in 1917, a major aim of the cosmologists has been to determine the curvature of the space that we and everything else live in. Is the universe closed and therefore finite, as Einstein originally believed? Or is it flat, as the consensus view presently is, or is it perhaps negatively curved? Questions like these could only be asked after the recognition, in the first half of the nineteenth century, that geometries other than the Euclidean are possible. While the development of non-Euclidean geometry is well researched by historians of science, this is not the case when it comes to the astronomical and physical aspects in the pre-Einsteinian era. The aim of this essay is to remedy the situation, to some extent, to take a closer and more systematic look at how people in the period from about 1830 to 1910 thought about curved physical and astronomical space. My focus is on cosmic space in particular, the essay being intended as a contribution to the history of cosmology no less than to the history of geometrical thought.

As noted by Jürgen Renn and Matthias Schemmel, astronomy might have contributed more than it actually did to the emergence of the theory of general relativity.[1] Among the reasons for its limited impact on the development was that few astronomers took the possibility of curved space seriously, and also that cosmology, in the sense of the science of the universe at large, was considered a somewhat disreputable area peripheral to astronomy proper. This again reflected the positivist attitude prevalent in the astronomical community in the period from about 1850 to Einstein's breakthrough in 1915.

I start with a brief and general review of the early phase, from Gauss in the 1810s to Helmholtz in the 1860s. During this half-century, the contributions of Lobachevsky and Riemann stand out as the most relevant, although Riemann was not concerned with the astronomical consequences of his theory. The sections 3 to 5 deal with the German astrophysicist Karl F. Zöllner and his remarkable, and somewhat overlooked, use of Riemannian geometry in a truly cosmological context. Only few astronomers were seriously concerned with the possibility of curved space, and correspondingly few mathematicians expressed interest in the astronomical aspects of non-Euclidean geometry. The Briton Robert S. Ball, the American Charles S. Peirce, the Frenchman Paul Barbarin, and the Germans Karl Schwarzschild and Paul Harzer were the most notable of the small crowd. Finally, Section 8 briefly looks at the situation about 1910 and how the curved-space universe a few years later entered as a solution of Einstein's relativistic field equations, and thereby revolutionized cosmology. On the other hand, Einstein's theory did not provide an answer to the old question of whether space is curved or not.

---

[1] Renn and Schemmel 2012, p. 8. See also Schemmel 2005.

## 2.  Background: Non-Euclidean geometries of space

As a classical case in the history of mathematical thought, the emergence and early development of non-Euclidean geometry has been thoroughly investigated by historians and mathematicians.[2] It is generally agreed that the eminent mathematician, physicist and astronomer Karl Friedrich Gauss was the first to conceive the possibility of a non-Euclidean geometry, although he refrained from publishing anything about it and did not study the possibility systematically. However, as early as 1817 – a century before Einstein's theory of the closed universe – he had arrived at the conclusion that the ordinary,

---

[2]  Among the numerous works are Bonola 1955, Rosenfeld 1988, Gray 1989, and Gray 2007. The popularity of non-Euclidean geometry in nineteenth-century mathematics is illustrated by Sommerville 1911, a bibliography which includes about 4,000 titles. By comparison, the earlier bibliography Halsted 1878 mentions only 174 titles.

Euclidean geometry was not true by necessity. This was a view that flatly contradicted the widely accepted Kantian epistemology according to which space, far from being empirical, was a transcendental property (in Kant's sense) that conditioned the very possibility of empirical knowledge. As Gauss wrote in a letter to his correspondent, the Bremen astronomer Heinrich Wilhelm Olbers: "Maybe in another life we shall attain insights into the essence of space which are now beyond our reach. Until then we should class geometry not with arithmetic, which stands purely a priori, but, say, with mechanics."[3]

Gauss' interest in a geometry different from Euclid's was increased by his reading in 1818 of a manuscript by Ferdinand Karl Schweikart, an amateur mathematician and professor of law. In Schweikart's *Astralgeometrie*, the sum of angles in a triangle was less than two right angles. He may have called it astral geometry to suggest that the system might be valid on an astronomical scale, such as he indicated in his brief memorandum.[4] The same year Gauss, who at the time served as director of the Göttingen Observatory, was asked to undertake a major cartographic survey project with the purpose of mapping the state of Hanover (to which Göttingen belonged) by means of triangulation. As part of this project he made measurements of unprecedented precision of a triangle extending between three mountain peaks. The sides of the Brocken-Hohehagen-Inselsberg triangle were approximately 69, 85 and 107 km. For a long time it was generally believed that the theoretical purpose of these measurements was to test the assumption of Euclidean geometry, namely, to establish whether or not the sum of the angles in the triangle deviated from 180°. However, he never actually said so, and today most historians agree that

---

[3] Letter of 28 April 1817, in Gauss 1900, p. 177. For details about Gauss' pioneering contributions to non-Euclidean geometry, see Reichardt 1976.
[4] Bonola 1955, pp. 75-77. Gray 2007, pp. 91-94.

Gauss' work had nothing to do with the possibility of physical space being non-Euclidean.[5]

Gauss realized that if the question could be settled at all, it would require astronomical measurements over much larger, stellar distances. Although he seems to have been aware that information about the curvature of space could in principle be obtained from data of stellar parallaxes, he did not pursue this line of reasoning.[6] In response to a letter from Gauss, the German-Baltic astronomer Friedrich Wilhelm Bessel admitted that "our geometry is incomplete and should be supplied with a hypothetical correction that disappears in the case that the sum of angles in a plane triangle = 180°." He continued: "This would be the *true* geometry, whereas the Euclidean is the *practical*, at least for figures on the earth."[7]

Apart from Gauss' anticipations, the founders of non-Euclidean geometry were unquestionably the Hungarian mathematician János Bolyai and the Russian Nikolai Ivanovich Lobachevsky, both of whom (contrary to Gauss) published their independent discoveries that a geometry different from and as valid as Euclid's is possible. While Bolyai's sole work on what he called "absolute geometry" dates from 1831, Lobachevsky's first study was from 1829,[8] followed in 1840 by a booklet in German, *Geometrischen Untersuchungen zur Theorie der Parallellinien*. Although working independently and in almost complete isolation, the two theories were remarkably similar. Both mathematicians believed that the truth of Euclidean geometry was a question to be determined empirically, but it was only the ten years older Lobachevsky, at the University of Kasan, who seriously contemplated the problem from the perspective of his new geometry. He suspected that the truth of geometry, Euclidean or not, "can only be verified, like all other laws of nature, by experiment, such as astronomical observations."[9] As a young student at Kasan University, Lobachevsky had studied astronomy under the Austrian Johann Joseph Littrow, who in 1810 had established an observatory at the university. Littrow, who later became director of the Vienna Observatory, recognized the outstanding mathematical abilities of his student and also made some astronomical observations with him.

---

[5] The myth was first criticized in Miller 1972, with comments and discussion in *Isis* 65 (1974), 83-87, and subsequently, in much greater detail, in Breitenberger 1984. A few historians of science argue that Gauss' measurements may well have had the purpose of testing Euclidean geometry (Scholz 2004).

[6] See, e.g., Gauss to Heinrich C. Schumacher, 12 July 1831, in Peters 1860, pp. 268-271.

[7] Bessel to Gauss, 10 February 1829, in Gaus and Bessel 1880, p. 493.

[8] Lobachevsky read his first paper on the principles of geometry on 12 February 1826 at a meeting in Kasan, but the paper was never published.

Already in his 1829 paper in the *Kasan Messenger* Lobachevsky suggested that one consequence of his "imaginary" (or hyperbolic) geometry might be tested by astronomical means, namely, that the angle sum of a triangle is always less than 180° and the more so the bigger the triangle becomes. He reasoned that this prediction might be checked by considering the parallax of stars such as Eridan 29, Rigel and Sirius. Quoting a smallest parallax value of 1".24 for the latter star, he concluded that the angle sum of the triangle spanning the sun, the earth and Sirius deviated from the Euclidean value of 180° by at most 0".00000372. This tiny deviation strongly suggested that space was Euclidean, and yet Lobachevsky refrained from drawing the conclusion in firm terms.[10] Realizing that while it could in principle be proved that astronomical space is non-Euclidean, it could never be proved to be Euclidean, he tended to see his calculations as inconclusive, perhaps because the measured area of space had not been large enough. At any rate, at the time no reliable determination of a stellar parallax had yet been made (although many had been announced). Only in 1838 did Bessel succeed in finding an annual parallax of 0″.3136 for the star 61 Cygni, corresponding to a distance from the earth of 657,000 astronomical units (1 AU $\cong 1.5 \times 10^8$ km). The modern value of the parallax of Sirius is 0″.37, less than a third of the value adopted by Lobachevsky. In another line of reasoning Lobachevsky showed that, if the world geometry is hyperbolic, the curvature constant must be greater than 1.66 $\times 10^{-5}$ AU.

---

[9] Lobachevsky, "Neue Anfangsgründe der Geometrie," originally published in Russian in 1835. Here from the German translation in Engel 1898, pp. 67-235, on p. 67. On Lobachevsky's life and career, see Vucinich 1962, a study that stresses the philosophical roots of Lobachevsky's geometry while it ignores his attempts to relate it to astronomical problems.

[10] Daniels 1975. See also Bonola 1955, pp. 94-96.

Lobachevsky also discussed the relevance of his new geometry to physical and astronomical space in later publications, such as his *Pangeometry* published in French in 1856, the year of his death.[11] In this work he argued that, assuming space to be hyperbolic, there must be a minimum parallax for all stars irrespective of their distances from the earth. (In Euclidean space, the parallax tends toward zero as the distance increases toward infinity.) His general conclusion was that since the deviation from flat space was smaller than the errors of observation, Euclidean geometry was a perfect approximation for all practical purposes. For the next sixty years, no one questioned the conclusion.

The ideas of non-Euclidean geometry pioneered by Gauss, Lobachevsky and Bolyai circulated painfully slowly in the mathematical community. It was only when they were taken up and presented in a better argued and clearer way by the Italian mathematician Eugenio Beltrami in 1868 that non-Euclidean geometries truly entered the world of mathematics and then resulted in a revolution in geometry. As illustrated by the number of titles on non-Euclidean geometry in Duncan Sommerville's bibliography, from about

---

[11] Lobachevsky 2010. *Pangeometry* first appeared in a Russian publication of 1855.

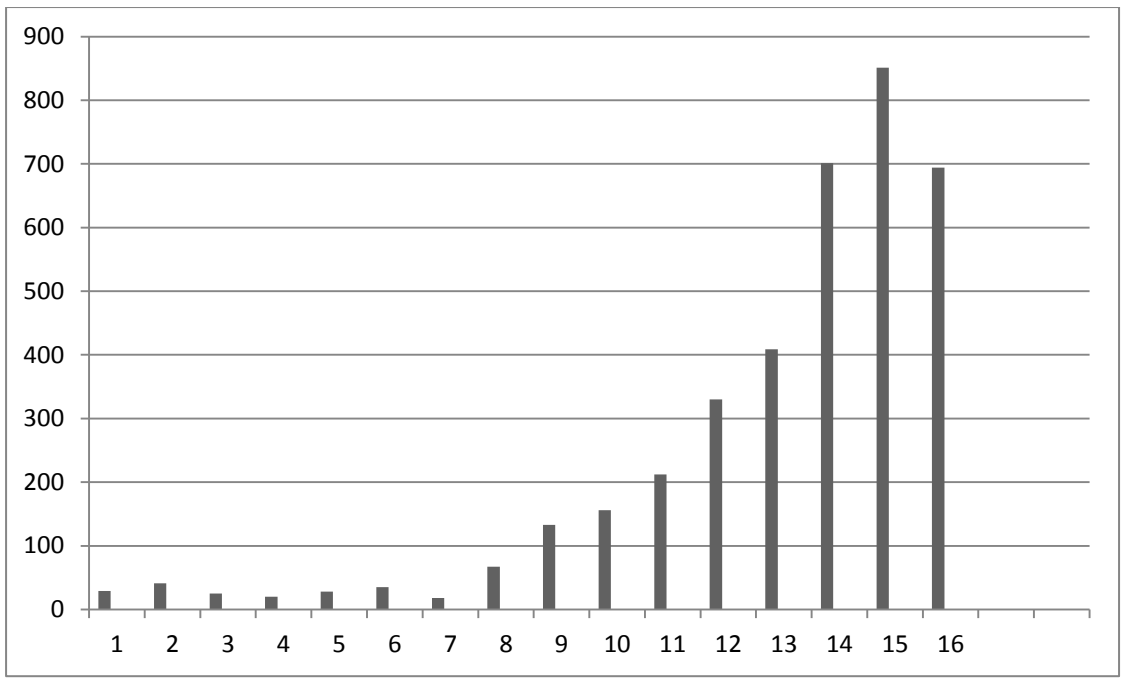

The diagram shows the number of titles per 5-year interval in Sommerville's 1911 bibliography on non-Euclidean geometry, starting with 1830-1834 and ending with 1905-1909. Of the 4,016 titles listed in the bibliography, by far most were written in German (1,149 or 28.6%), French (884 or 22.0%), Italian (848 or 21.1%), and English (723 or 18.0%).

1870 interest in the new geometries experienced a boom. With respect to the physical and astronomical sciences, the ideas took even longer to catch on, and it was only with the general theory of relativity they became a natural component of these sciences.

In 1854 the young Göttingen mathematician and physicist Bernhard Riemann put the concept of curvature as an intrinsic property of space on a firmer basis. His lecture read in Göttingen on 10 June 1854 was a *Habilitationsvortrag* (qualifying lecture) the subject of which – "On the Hypotheses on which Geometry is Founded" – was chosen by Gauss. It remained unpublished until 1867 and even then did not attract much attention, although in Germany Riemann's ideas became known through popular lectures by Hermann von Helmholtz. It was only after the British mathematician

William Kingdon Clifford translated the lecture into English in 1873 that it came to be seen as a visionary address on the possible geometrization of physics. Riemann's thoughts also became known to the English-speaking world through articles by Helmholtz that appeared in *Mind*, a new quarterly journal for philosophy and psychology founded in 1876.[12]

Clifford had introduced Riemann's ideas to a British audience three years earlier, in a communication to the Cambridge Philosophical Society of 21 February 1870 which however remained unpublished until 1876. "Riemann has shewn," he wrote, "that there are different kinds of space of three dimensions; and that we can only find out by experience to which of these kinds the space in which we live belongs."[13] Moreover, he extended Riemann's ideas by speculating that physical phenomena could be fully reduced to properties of space curvature varying between one portion of space to another. Heat, light and magnetism might be mere names for tiny variations in the curvature of space, he boldly hypothesized. Also in an address to the 1872 meeting of the British Association for the Advancement of Science, Clifford pointed out that within the framework of Riemannian geometry the association between limitedness and finite extent was invalid. He emphasized, such as earlier contributors to non-Euclidean geometry had done, that the geometrical structure of space was a question of empirical facts and not of metaphysics.[14]

---

[12] Helmholtz 1876. Helmholtz came to his ideas about non-Euclidean geometry more or less independently of Riemann, whose lecture he only became aware of in the spring of 1868. Riemann's theory of constant-curvature geometry was more general than Helmholtz's, which was restricted to three dimensions. For accounts of Helmholtz's view of geometry, see DiSalle 1993 and Carrier 1994.

[13] Clifford 1876, reprinted in Gray 1989, p. 225. On Clifford's brief and enigmatic paper, see Farwell and Knee 1990, who argue that Clifford rather than Riemann should be credited as having anticipated the Einsteinian concept of a geometrization of physics.

[14] Clifford 1947, p. 11.

Riemann's lecture of 1854 was intended for a non-specialized university audience and not for mathematicians, and perhaps for this reason it was neither very clear nor very specific. It should also be noticed that non-Euclidean geometry was *not* the main subject of his qualifying lecture and only appeared by implication. In fact, Riemann did not mention the term at all and also did not refer to either Bolyai or Lobachevsky.[15] Nonetheless, he probably knew Lobachevsky's theory from an article in French that the Russian mathematician had published in 1837 in Crelle's *Journal für reine und angewandte Mathematik*. At any rate, the Göttingen address of 1854 eventually became a classic in the history of non-Euclidean geometry.[16]

Riemann pointed out that while Euclidean space was characterized by the particular metric

$$ds^2 = dx^2 + dy^2 + dz^2 ,$$

there were other metrics that generated alternative spaces. Importantly, his address led to the standard distinction between three geometries of constant curvature, corresponding to flat or Euclidean space (curvature constant $k = 0$), spherical space ($k = +1$), and hyperbolic space ($k = -1$). Although there is any number of possible geometries, only these three have properties that make them candidates for the real space: they are homogeneous and isotropic, and also invariant under rotation and translation. Riemann was the first to point out

[15] It may not be irrelevant to mention that one of the panellists evaluating Riemann's lecture was the philosopher Hermann Lotze, who much disliked non-Euclidean geometry. He maintained that the axioms of Euclidean geometry were self-evident truths. For Lotze's arguments against non-Euclidean geometry, see Russell 1897, pp. 93-115.

[16] On Riemann and his contributions to geometry, see Gray 2007, pp. 187-201, which includes the major part of the 1854 lecture. See also Torretti 1978, pp. 82-106, and Farwell and Knee 1990. After Gauss' death in 1855, his chair in Göttingen was taken over by Gustav Lejeune Dirichlet, and in 1859 Dirichlet was succeeded by Riemann.

that, in the case of constant positive curvature, the traditional identification of a
finite three-dimensional space with a bounded space is unwarranted:

> In the extension of space-construction to the infinitely great, we must
> distinguish between *unboundedness* and *infinite extent*, the former belongs
> to the extent relations, the latter to the measure-relations. … The
> unboundedness of space possesses in this way a greater empirical
> certainty than any external experience. But its infinite extent by no means
> follows from this; on the other hand, if we assume independence of bodies
> from position, and therefore ascribe to space constant curvature, it must
> necessarily be finite provided this curvature has ever so small a positive
> value.[17]

Riemann only referred to astronomical questions in passing, possibly an
allusion to Lobachevsky: "If we suppose that bodies exist independently of
position, the curvature is everywhere constant, and it then results from
astronomical measurements that it cannot be different from zero; or at any rate
its reciprocal must be an area in comparison with which the range of our
telescopes may be neglected." On the other hand, in microphysics, where space
need not be isotropic, "the curvature at each point may have an arbitrary value
in three directions, provided that the total curvature of every measurable
portion of space does not differ sensibly from zero." Therefore, "we are … quite
at liberty to suppose that the metric relations of space in the infinitesimal small
do not conform to the hypotheses of geometry." Although Riemann accepted
the idea of just one physical space, he did not accept that the geometry of this
space could be known with certainty. Our experiences about the physical world
could be consistent with geometries of many different kinds.

Elsewhere in his lecture Riemann vaguely and somewhat enigmatically
suggested that the metrical structure of space might be causally connected to the state and configuration of matter particles. He left open the possibility that on a microphysical scale, say between atoms or molecules, the curvature of space might vary, if in such a way that the averaged curvature over measurable distances becomes zero or unappreciably close to zero. Nonetheless, for him physical events took place in the arena of space and were not (as Clifford would suggest) parts of or manifestations of space. Geometry and physics remained separate.[18]

---

[17]  Riemann 1873, p. 36. The other quotations are from the same source.

As mentioned, Riemann's famous address is often seen in the context of non-Euclidean geometry, but this is to some extent a later rationalization. It is perhaps better placed in a larger context that relates to Riemann's physical ideas and in particular to his dynamical conception of matter, forces and ether inspired by natural philosophy (as distinct from the *Naturphilosophie* tradition).[19] Apart from being a brilliant mathematician, Riemann was also deeply engaged in electrodynamics in the tradition of Wilhelm Eduard Weber, Gauss' close collaborator. For example, in a paper of 1858 he proposed a new theory of electromagnetism in which the electric force propagated in time and he later developed it into an electric theory of light different from Maxwell's field theory. Riemann's ambition was nothing less than a total theory of physics, including electricity, magnetism, gravity and light, based on a single mathematical law.[20] However, he did not take an interest in the space of the astronomers. Questions about the global properties of space he cut short as "idle questions."

---

[18]  I rely on the interpretation in Farwell and Knee 1990.

[19]  Bottazzini and Tazzioli 1995. See also Boi 1994, who relates Riemann's ideas to the much later ideas of cosmological models based on the general theory of relativity.

[20]  On Riemann as a physicist and a representative of the mathematics-physics relationship in mid-nineteenth century Germany, see Jungnickel and McCormmach 1986, pp. 174-181.

## 3. Astrophysicist, controversialist, spiritualist

Although Riemann's emphasis on the possibility of an unbounded yet finite space implicitly addressed an old cosmological conundrum, it failed to attract interest among astronomers. It took nearly two decades before a scientist made astronomical use of Riemann's insight. The 1872 contribution of the Leipzig astrophysicist Johann Karl Friedrich Zöllner is not well known, but it deserves more than a footnote in the history of non-Euclidean physical geometry. Before I discuss Zöllner's work and its significance, a sketch of his life and career will be useful.[21]

After studies at the University of Berlin, in 1857 Zöllner went to Basel, where he wrote his doctoral dissertation on photometry, an area of physics in which he soon became an authority. A skilled experimentalist and designer of instruments, he invented an astrophotometer to measure the feeble light from celestial bodies, including planets, comets and stars.[22] In 1862 he moved to Leipzig, where he continued his work in astrophotometry and in 1866 was appointed professor. He established an astrophysical research programme, the first of its kind, and emerged as one of the leaders of the new, interdisciplinary field of astrophysics. Indeed, he may have been the first to use the name "astrophysics," which he introduced in 1865. Apart from his fundamental work in astrophotometry, he took up the new and exciting field of astrospectroscopy, in particular by constructing a so-called reversion spectroscope.

---

[21]  The scholarly literature on Zöllner is limited and mostly in German. Hermann 1982, published on the occasion of the centenary of Zöllner's death, provides a useful overview coloured by being written in the German Democratic Republic. (It presents Zöllner as a dialectical materialist and includes many references to Marx, Engels and Lenin). Koerber 1899 includes selections of Zöllner's correspondence, and Meinel 1991 places Zöllner in the general context of German science and culture.

[22]  On the importance of Zöllner's photometer and its use in astronomy, see Pannekoek 1961, pp. 385-387.

In addition to his experimental work, Zöllner also made important studies of theoretical problems in astronomy and physics. These included electrodynamics, solar theory, sunspots, stellar evolution, and the theory of comets. In his *Natur der Cometen* from 1872 he developed an electrical theory of comets that for a period was widely admired. According to Zöllner, when a comet approached the sun, slow evaporation of the comet's nucleus would create intense electrical phenomena that were responsible for the formation of the cometary tail. Whereas the traditional view was that the light from the comets was due to incandescent cometary particles, Zöllner's theory explained the light as caused by electrical disturbances.[23]

Zöllner was a tireless advocate of the theory of electrodynamics developed by Weber, who had been professor of physics in Leipzig 1843-1849 before he returned to Göttingen. Weber's ambitious theory was based on a fundamental force law acting between hypothetical charged particles, and it assumed the interaction between the particles to occur instantaneously.[24] Not only did Zöllner admire and accept Weber's force law and associated atomistic theory, he also argued that it was of universal significance and valid for all terrestrial and cosmic phenomena. Weber's law, he suggested, could be translated into a law of gravitation superior to Newton's law of attraction, in the sense that the latter was merely a special case of Weber's.

In Zöllner's extended version of Weber's theory, the interaction between two charged particles of opposite sign differed slightly from the interaction between two particles of the same sign. For two elementary particles of mass $m$ and charges $\pm e$ he assumed the attractive electrical force to exceed

---

[23] For theories of comets, including Zöllner's, see Heidarzadeh 2009, especially pp. 224-228.

[24] See Jungnickel and McCormmach 1986, pp. 139-152.

the repulsive force by a factor $(1 + \gamma)$. For the tiny quantity $\gamma$ he obtained the expression

$$\frac{1}{2\gamma} = \frac{e^2}{Gm^2} \cong 3 \times 10^{40} \, ,$$

from which $\gamma = 1.7 \times 10^{-40}$.[25] Thus, the electric interaction between two neutral bodies would not be zero. A very small residual force would remain between them, and this residual electric force he identified with the gravitational attraction.[26] Zöllner's electro-gravitational force of attraction had the same form as Weber's electrical law, including that it contained a velocity-dependent term. Among other things, in collaboration with the Leipzig mathematician Wilhelm Scheibner he used the theory in an attempt to explain the anomalous motion of Mercury's perihelion, one of the major problems in astronomy until it was finally solved by Einstein. He did not succeed.

*Natur der Cometen* was a remarkable work in more than one sense. The major part of the 600-page long book was not about comets, but instead a strange mixture of history and philosophy of science and unconstrained, chauvinistic charges of plagiarism. Zöllner's main targets were British scientists, including luminaries such as William Thomson, Peter G. Tait and Charles Darwin, but he also attacked Helmholtz, Emil du Bois-Reymond and the chemist August Wilhelm Hoffmann, three of the most powerful men in German science. The book aroused a storm of controversy, not only in Germany but also in England, and had the effect that Zöllner became increasingly marginalized as a scientist, almost a *persona non grata*.[27] It did not help his reputation when he turned to spiritualism, which he did about 1877, after he had met William Crookes in England two years earlier. What fascinated him was not only the explanation of Crookes' sensational light-mill or radiometer – perhaps demonstrating "a fourth state of matter" – but also the English chemist's deep interest in spiritualism.[28]

---

[25] Zöllner 1876, p. xii. This may have been the first recognition of the later so famous ratio between the gravitational and the electromagnetic interaction. The dimensionless number $F_{grav}/F_{el} \sim 10^{-40}$ is often referred to as either Weyl's or Eddington's number. It remains unexplained to this day. See, for example, Treder 1981.

[26] Zöllner 1872, pp. 333-335. Zöllner 1882. This kind of theory was sometimes known as the "Zöllner-Mossotti-Weber theory," the middle name being a reference to the Italian physicist Ottaviano Fabrizio Mossotti, who as early as 1836 had proposed somewhat similar ideas. The Zöllner-Mossotti-Weber theory was taken fairly seriously and developed or modified by several physicists. For example, as late as 1900 H. A. Lorentz derived, on the basis of his Maxwellian electron theory, a gravitational law of a form similar to the Zöllner-Weber law.

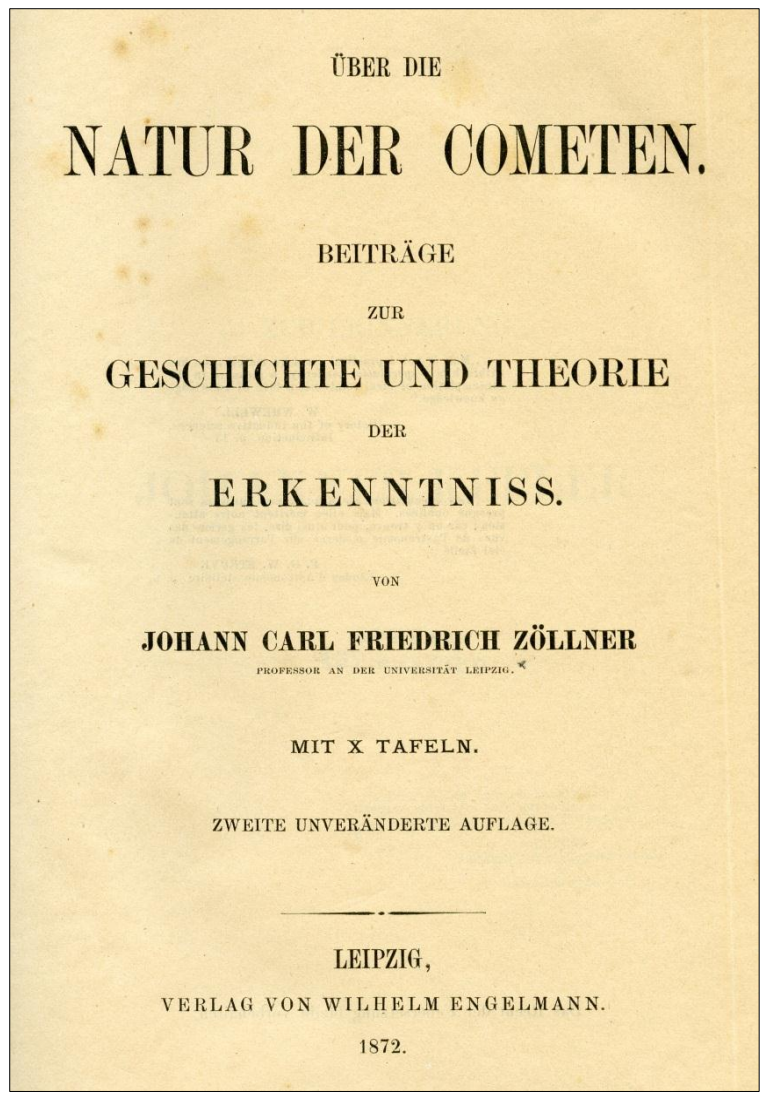

---

[27] One may get a taste of the controversy from Tait 1878, an acerbic review in which Zöllner is ridiculed as a raving metaphysician. Some of his critics suggested that he was mentally ill. On the Zöllner *cause celèbre* and its consequences, see also Small 2001, pp. 59-80, a slightly different version of which appeared as Small 1994. On Helmholtz's response to Zöllner's book, see Cahan 1994.

Soon convinced of the reality behind spiritualist manifestations, Zöllner investigated them in great detail, attempting to integrate the spirits with both Weberian physics and his own, highly unorthodox version of Christian theology. In 1878 he published a paper "On Space of Four Dimensions" in *Quarterly Journal of Science*, a journal edited by Crookes, in which he reported his experiments with the astonishing American medium "Dr." Henry Slade.[29] Among the major results of Zöllner's efforts in this area of unconventional research was an elaborate *Transcendentale Physik* published in 1878 (with an English translation of 1880) and a treatise of 1881 with the title *Naturwissenschaft und Christliche Offenbarung* in which he explained the miracles of the Bible in terms of four-dimensional space. To his mind, the project of a transcendental physics including both material and spiritual phenomena was but a natural extension of the astrophysical project of accommodating terrestrial and celestial phenomena within the same theoretical framework.

Not satisfied with simply accepting the spirits of deceased persons, as they appeared in séances, Zöllner argued that they were visitors from a hidden fourth dimension of space which he identified with Kant's absolute space, a reality in its own right and independent of the existence of matter. During the last decades of the nineteenth century beliefs of this kind were widespread, and

---

[28]  Zöllner constructed his own version of Crookes' instrument, which was sold as the "Zöllner radiometer." On Crookes' and other British Victorian scientists' interest in spiritualism, see Oppenheim 1985 (where the radiometer is dealt with on pp. 352-354).
[29]  Zöllner 1878. Treitel 2004, pp. 4-7.

Zöllner only took them more seriously than most.[30] It was sometimes contended that if our space is curved, it must be contained in a flat space of a higher dimension, in the same way that a two-dimensional space is embedded in our three-dimensional space. Although four-dimensional "hyperspace" was often mixed up with ideas of non-Euclidean geometry in this or other ways, in reality there is no necessary connection between them. Clifford dismissed the connection as groundless, such as did later mathematicians.[31] A curved space does not need to be curved "in" another space. As already Gauss was aware of, curvature is an intrinsic property of space, and it can be determined by measurement made solely within that space. Zöllner's belief in a spiritual fourth dimension received inspiration from his knowledge of non-Euclidean Geometry, which he sometimes used for purposes of illustration, but it did not depend on it.

Nor did his claim of a fourth dimension rely exclusively on his belief in a spiritual world, for he held the claim even before his conversion to spiritualism. In a book of 1876 he argued that a fourth dimension was needed for epistemological reasons, in order to understand the symmetry between three-dimensional objects, such as left- and right-handed gloves. The phenomenal objects in our three-dimensional world must be "projections of objects in a space of four dimensions."[32] He considered the insight to be of revolutionary importance to science as it heralded a change in the world view on a scale comparable to the one Copernicus had initiated. As a result of the Copernican revolution, he wrote, "the third dimension of the firmament turned into a key that unlocked [the secrets of] the celestial motions, and in the future the assumption of a fourth dimension will provide the key for explaining all phenomena in three-dimensional space."[33] Zöllner was not the only one to describe the non-Euclidean geometry in such terms. "What Copernicus was to Ptolemy, that was Lobatshcewsky to Euclid," Clifford said.[34]

[30] For the late-nineteenth century fascination of the fourth dimension among both mathematicians, physicists and the lay public, see Bork 1964 and Beichler 1988. Sommerville 1911 mentions a large number of titles on four-dimensional spaces, including a group of "the realm of the spirits." In Germany, Helmholtz was instrumental in popularizing four-dimensional hyperspace, but without interpreting it spiritually. Although Zöllner and Helmholtz shared an empiricist view on space perception, the consequences they drew from it were very different (Stromberg 1989). On Zöllner's transcendental physics and spiritualism in Germany, see Treitel 2004, pp. 3-17, and, for a contemporary description, Wirtz 1882. The reception of Zöllner's speculations in British spiritualist circles is examined in Valente 2008.

[31] Farwell and Knee 1990, p. 112. For a more elaborate refutation, see, for example, Russell 1897, pp. 101-108.

It should be added that although much of the last decade of Zöllner's life was occupied with spiritualism, philosophical speculations and political writings (some of it of a strongly xenophobic and anti-Semitic nature), he continued doing scientific work of a more conventional, non-transcendental kind. Thus, he developed a theory of the origin of the earth's magnetism according to which the magnetism was due to electrical currents in the fluid core of the earth.

Later in the century speculations of four-dimensional space became popular, almost fashionable, if in most cases outside science. The English-American mathematician and author Charles Howard Hinton wrote a series of articles and books in which he claimed that the fourth dimension might explain physical phenomena such as the nature of ether and electricity. His project had points of similarity to Zöllner's, but he seems to have been unaware of his German predecessor and did not endow the extra dimension with spiritual powers. Although Hinton included non-Euclidean geometry in his writings, he realized that the ideas of Bolyai and Lobachevsky were "something totally different from the conception of a higher space." On the other hand, he judged, undoubtedly correctly, that "metageometry had a great influence in bringing the higher space to the front as a working hypothesis."[35] The speculations of Hinton and other writers in the same tradition did not refer to the possible curvature of cosmic space, for which reason they are peripheral to the present study.

---

[32] Zöllner 1876, p. lxxix.

[33] Ibid. Ptolemaic astronomy was essentially two-dimensional, while the Copernican version allowed astronomers to determine the distances of the planets and in this sense to provide the universe with a depth-dimension.

[34] Quoted in Vucinich 1962, p. 469.

## 4. Zöllner's closed universe

During the nineteenth century problems of cosmology were of limited interest to astronomy, a science which primarily dealt with the solar system, the stars making up the Milky Way, and the enigmatic nebulae. In the spirit of positivism, the large majority of astronomers tended to conceive the universe at large as a field for philosophical study rather than scientific exploration. As long as the riddle of the nebulae – that is, whether or not the nebulae belonged to the Milky Way – remained unsolved, cosmology was bound to remain speculative, hence unscientific. "As there are no observed facts as to what exists beyond the farthest stars, the mind of the astronomer is a complete blank on the subject," the American astronomer Simon Newcomb admitted in a review of 1907. He added, "Popular imagination can fill the blank as it pleases."[36]

---

[35]  Hinton 1912, pp. 57-59 (first published 1904). On Hinton and the fourth dimension, see Bork 1964 and Beichler 1988. Hinton published his first paper on fourth-dimensional space in 1880, at a time when Zöllner was still alive. The best known work on higher-dimensional space in the Victorian era was the English schoolmaster and theologian Edwin A. Abbott's novel *Flatland* of 1884.

[36]  Newcomb 1907, p. 362. The same attitude permeated Agnes Clerke's masterly study of nineteenth-century astronomy, a work which had little to say about cosmology and

Although astronomers rarely dealt with the universe as a whole, there was in the last half of the nineteenth century a lively discussion of issues of a cosmological nature, in particular the age-old question of the spatial and material extension of the universe: is it finite or infinite? (There was a corresponding discussion of the temporal extension.) In agreement with the generally held view, Richard Proctor, a popularizer of astronomy, phrased the central issue as follows: "The only question for us is between an infinity of occupied space and an infinity of vacant space surrounding a finite [material] universe."[37] The question of infinity turned up in three, to some extent interrelated problems, namely, Olbers' optical paradox, the gravitational paradox, and the paradox of entropy increase. Of these, it is the first one that is of most interest in the present context.

This is not the place to discuss the famous paradox that traditionally is associated with the name of Olbers but which can be traced back to the days of Kepler.[38] Olbers' attempt to dissolve the paradox of the dark night sky dates from 1823, although his paper was only published three years later. Accepting an infinity of stars, he argued that interstellar space was not perfectly transparent but filled with a tenuous medium that over long distances would diminish the intensity of starlight. Olbers' assumption of interstellar absorption was generally accepted by nineteenth-century astronomers who consequently did not see the paradox as truly paradoxical. This only changed in the early part of the new century, when it was realized that space is much more transparent nothing about the possibility of non-Euclidean space (Clerke 1886). For an account of cosmology in the period 1840-1910, see Kragh 2007, pp. 83-125.

[37] Quoted in Jaki 1969, p. 183.

[38] Detailed expositions of Olbers' paradox are given in Jaki 1969 and Harrison 1987, of which the first source is particularly valuable from a historical point of view, and the latter from a scientific point of view. Harrison 1987, p. 173, refers to Zöllner's solution of Olbers' paradox in terms of closed space, but mistakenly dates it to 1883.

than assumed by Olbers and his followers. There were other proposals of defusing the paradox: one of them was to assume a non-homogeneous (hierarchic) distribution of stars, and another was to postulate that the stars had not existed for ever. A third possibility was to assume that the stellar universe was finite, but this hypothesis was widely considered unattractive, both for philosophical reasons and because of the gravitational collapse problem that already Newton had made aware of. A fourth possibility remained unmentioned until the 1930s, namely, that cosmic space is in a state of expansion. The full explanation of Olbers' famous paradox is complex and belongs to the twentieth century.

Zöllner's *Natur der Cometen* included a 16-page chapter on "The Finitude of Matter in Infinite Space" (*Ueber die Endlichkeit der Materie im unendlichen Raume*) in which he offered an original solution to Olbers' paradox in terms of a universe of constant positive curvature. However, Zöllner's aim was more than just demonstrating how an astronomical problem could be explained on the basis of this hypothesis; it was a general argument for a Riemannian cosmic space, with the solution of Olbers' paradox appearing merely as one consequence of the argument.[39] In a publication of 1871 dealing with the stability of cosmic matter he had considered a very large yet finite space, but at the time without connecting it to Riemann's geometry.[40] This is what he did in his essay of 1872. In his systematic discussion of the finite versus the infinite in the universe, he assumed, for the sake of discussion, that there is only a finite amount of matter in the world. He then listed some general assumptions of cosmic matter according to Newtonian physics:

1. The quantity of matter is located in an unbounded Euclidean space.
2. The time during which matter is located in this space is infinite.
3. Apart from its usual properties, cosmic matter tends to evaporate at any temperature above absolute zero.

---

[39]  Accounts of Zöllner's arguments can be found in Jaki 1969, pp. 158-161, Small 2001, p. 65-67, and Kragh 2008, pp. 158-160. However, they are not generally known, and in the history of science and ideas Zöllner often appears, if at all, as just a crackpot scientist advocating a spiritual fourth dimension. See, for example, Jammer 1993, pp. 181-182, and Scriba and Schreiber 2005, p. 427.
[40]  Zöllner 1871.

Zöllner now argued that in an unbounded (and therefore infinite) Euclidean space any finite amount of matter would dissolve to zero in an infinity of time. Given the actual existence of matter of non-zero density, at least one of the assumptions 1 and 2 must be wrong. As to assumption 2, he realized that a universe of finite age would solve the problem of the vanishing matter density. One could imagine, he wrote,

> an act of creation in which had begun, at a time in the finite past, a certain finite initial state of the world, which now continues in a way that is imperceptible to our senses and periods of time [and which] approaches the several times mentioned end state in which, after an infinite time, the elements of matter are to be found in infinitely large distances. From a physical point of view, such a process would be equivalent with a gradual dissolution of the world into nothing or with the annihilation of the world.[41]

However, Zöllner was unwilling to accept a limitation of time "either in the past, in the form of a definite beginning, or in the future, in the form of a definite end of all material changes." Not only would a beginning limit the causal chain arbitrarily, it would also contradict the Leibnizian principle of sufficient reason: there can be given no reason why the universe came into being a certain time ago rather than at any other time. He was therefore left with assumption 2, that cosmic space is Euclidean.

---

[41] Zöllner 1872, p. 306. This was Zöllner's only (and indirect) reference to the heat death and the entropic creation argument based on the second law of thermodynamics. For the discussion of these ideas, see Kragh 2008.

Zöllner was obviously well acquainted with the literature on non-Euclidean geometry. Apart from references to Gauss, Bolyai, Lobachevsky, Riemann and Helmholtz, he also cited recent works by Felix Klein and Jakob Rosanes.[42] He found Riemann's theory to be particularly valuable and quoted lengthy passages of the 1854 lecture. Zöllner had his scientific heroes as he had his scientific enemies, and among his heroes Riemann ranked almost as high as Weber and Kant.[43] As he saw it, Riemann's perspective was more important than that of the other mathematicians, for "it opens up for the deepest and most fruitful speculations concerning the comprehensibility of the world."[44] The possibility of a space of constant positive curvature was no less than the key that would unravel the secrets of the universe and dissolve the problems of a materially finite universe: "It seems to me that any contradictions will disappear … if we ascribe to the constant curvature of space not the value zero but a positive value, however small. … The assumption of a positive value of the spatial curvature measure involves us in no way in contradictions with the phenomena of the experienced world if only its value is taken to be sufficiently small."[45]

In this way Zöllner made Olbers' paradox disappear, at least to his own satisfaction, without having to assume a limitation of either cosmic time or space. According to later knowledge, based on the gravitational effect on scattered starlight, the riddle of cosmic darkness remains even in a static curved space. This is interesting from a scientific point of view, but of course it is of no historical relevance.[46]

---

[42] Namely, Klein 1871 and Rosanes 1870.

[43] Zöllner was a great admirer of Kant's cosmology as laid down in *Allgemeine Naturgeschichte und Theorie des Himmels* from 1755, a work he did much to publicize. For example, Zöllner 1872 included a 57-page chapter on Kant and his contributions to science. Nowhere did he mention that curved space and the fourth dimension were incompatible with the great philosopher's conception of space. On Zöllner's fascination of Kant, see Hermann 1976.

[44] Zöllner 1872, p. 312.

[45] Ibid., p. 308.

Zöllner thought that in a Riemannian space all physical processes would occur cyclically, indeed that the universe itself would be cyclic, so that time would go on endlessly. "In such a space, the parts of a finite quantity of matter, moving apart with a finite speed, could never reach infinitely distant points," he wrote. "After finite intervals of time, … they would converge again, and in this way transform kinetic energy to potential energy, and by separation from potential to kinetic energy, as periodical as a pendulum."[47] He noted with satisfaction that energy conservation would apply to his finite material universe, whereas he did not comment on the apparent irreconcilability between a cyclic universe and the irreversibility inherent in the second law of thermodynamics. A few years earlier, in an address to the German Association of Natural Scientists and Physicians in Frankfurt am Main, Rudolf Clausius had emphasized that his formulation of the second law in terms of increased entropy ruled out the popular notion of a cyclic world.[48] Nonetheless, belief in such a pulsating (Euclidean) universe continued to be widespread throughout the century, especially among thinkers of a materialist and positivist inclination. Zöllner was the only one who associated the cyclic universe with the curvature of space, but unfortunately he did not elaborate. Whatever he meant, he did *not* anticipate the later idea of a curvature of space varying periodically in time, an idea first proposed by Alexander Friedmann in 1922 within the context of relativistic cosmology.

---

[46] "The sky at night is equally bright in the two universes [closed and open] if the stars have similar distributions. Curvature, oddly enough, has no effect on the radiation." Harrison 1987, p. 173.

[47] Zöllner 1872, pp. 308-309.

[48] For ideas of a cyclic universe ca. 1850-1920 and references to the literature, see Kragh 2008.

Whereas Zöllner did not comment on the consequences of the second law of thermodynamics, he did apply a thermodynamic argument against the hypothesis of an infinity of stars embedded in a space with the capacity of absorbing starlight. Because, the radiant heat from the stars would then heat up the interstellar medium until it reached a state of thermal equilibrium and itself became radiant. Ultimately, the radiation problem would merely shift from the stars to the space. Therefore, "also the considerations offered by Olbers lead us to the assumption of a finite amount of mass in the universe, an assumption which … can be maintained only by postulating a non-Euclidean space."[49]

Clearly inspired by Riemann, and happy to admit the inspiration, in *Natur der Cometen* Zöllner speculated that curved space was dynamically active, in the sense of determining the laws of nature. For example, the law of inertia would have to be modified if space were curved.[50] Not even the divine force law of Weber was true a priori but somehow of cosmological origin, a speculation that bears some similarity to later versions of Mach's principle. And Zöllner went further than Riemann: whereas the Göttingen mathematician had declared that physics represented the "domain of another science," the Leipzig astrophysicist maintained that the science of the physical world belonged entirely to the field of Riemann's investigations. As mentioned, Zöllner was convinced that an extended form of Weber's force law was applicable also to gravitational physics and might, for example, explain the anomalous motion of

---

[49] Zöllner 1872, p. 311. As early as 1848, John Herschel had considered the same objection, but without drawing the conclusion that the universe must be finite (Jaki 1969, pp. 147-149).

[50] Zöllner 1872, p. 340. See also Leihkauf 1983. There is some similarity between the ideas of Zöllner and Clifford.

Mercury. He discussed the matter with his mathematics colleague in Leipzig, Wilhelm Scheibner, who developed a theory along this line.[51]

While Scheibner's theory, to which Zöllner referred in his *Natur der Cometen*, was based on Euclidean geometry, later in the century a few mathematicians attacked the problem by assuming space to be non-Euclidean. Wilhelm Killing and Carl Neumann derived orbits for Mercury moving in spherical space, in 1885 and 1886, respectively, and in 1902 Otto Liebmann did the same in the case of hyperbolic space.[52] This line of work, mathematical investigations of celestial mechanics in non-Euclidean space, continued even after Einstein had introduced his general theory of relativity and demonstrated that it solved the problem of the Mercury anomaly.[53]

## 5. Responses and criticism

Possibly because of the controversy caused by *Natur der Cometen*, Zöllner's new ideas of a fourth dimension and a closed universe became well known among readers in German-speaking Europe. In a letter of 1878, Otto Wilhelm Fiedler, a professor of mathematics in Zurich, informed him about recent developments in non-Euclidean geometry, a field he thought Zöllner had not sufficient insight in. Although critical, he praised Zöllner's "ideas about a connection between the structure of space and the sciences." In this context he referred to relevant work by Julius Plücker, professor of mathematics and physics in Bonn, and also to the Dublin astronomer Robert Ball (whom we shall meet in Section 6).[54]

[51] Zöllner 1872 (p. 334) mentioned Scheibner's ongoing work, which he only published much later (Scheibner 1897).

[52] On pre-relativistic attempts to explain the mercury anomaly, including those based on Weberian physics and non-Euclidean space, see Roseveare 1982, pp. 121-123, 162-165.

[53] Lense 1917.

The interest in Zöllner's speculations about cosmic space was mostly limited to philosophical circles, while astronomers and physicists paid almost no attention to them.[55] The only astronomer who referred to Zöllner's solution of Olbers' paradox may have been Wilhelm Meyer at the Geneva Observatory, who in a booklet of 1878 objected that it rested on a denial of "the most fundamental properties of space."[56] He did not specifically refer to Zöllner's space being Riemannian. As to the philosophers, Friedrich Nietzsche is known to have studied and appreciated Zöllner's book – he borrowed it from a library in 1872 and later bought a copy of it. He was generally in sympathy with the Leipzig professor, whom he in a letter described as an "honourable man," but it is unclear whether his sympathy extended to Zöllner's cosmological theory based upon Riemannian space.[57] Suffice to mention that some Nietzsche scholars have suggested that Nietzsche's philosophy of eternal recurrence

---

[54] Letter of 14 February 1878, in Koerber 1899, pp. 85-87. In Zurich, Fiedler taught geometry to, among others, Einstein and his friend Marcel Grossmann, who wrote his doctoral thesis under Fiedler. As well known, Einstein's development of the general theory of relativity relied crucially on Grossmann's mathematical expertise. In the history of mathematics, Plücker is known for his invention of line geometry and contributions to projective geometry; in the history of physics, for his pioneering work on discharges in rarefied gases, a line of research that eventually led to the discovery of the electron.

[55] Zöllner was not neglected in the astronomical literature, but his contribution to cosmology was. For example, while Clerke 1886 dealt with his photometric studies and theory of comets, she was silent about his idea of a finite universe. Strangely, so was Felix Koerber in his biography of 1899.

[56] Meyer 1878, p. 11.

[57] Small 2001, p. 62.

embodies principles of Riemannian geometry that he may have picked up in Zöllner.[58]

The same year that *Natur de Cometen* was published, it received a comprehensive critical commentary in a booklet written by Emil Budde, an author of scientific and cultural subjects and a privatdozent at the University of Bonn. After quoting a substantial part of Zöllner's essay on the finitude of space, Budde pointed out that the hypothesis disagreed with Clausius' law of entropy increase, which made him conclude that it was "absurd."[59] An advocate of the consensus view (at least among materialists) that space, time and matter are all infinite, he argued that Zöllner's universe failed to escape Olbers' paradox even though it contained only a finite number of stars. His discussion reveals that he did not realize that Zöllner's hypothesis rested on Riemann's notion of an unbounded yet finite space. According to Budde's misconception, at "the boundary of the world," which he likened to a huge mirror, the starlight must be reflected and thus destroy Zöllner's argument. He simply failed to understand that it is space itself which is closed, and that there is no boundary in the closed Zöllner-Riemann cosmos.

In 1875 the young philosopher Hans Vaihinger, later to become a recognized Kant scholar and the founder of "as-if" (*als-ob*) philosophy, gave a talk in Leipzig in which discussed the general scientific and philosophical problems associated with cosmology. Like many of his contemporaries, he focused on the heat death and its claimed corollary, a world with an origin in time. In this context he also dealt with Zöllner's arguments for a closed, non-

---

[58] Moles 1989, who argues that Nietzsche's doctrine of the eternal recurrence amounts to a cosmological hypothesis that makes sense only if understood on the background of Riemannian geometry. I consider his conclusion that "Zöllner's influence on Nietzsche is apparent" to be entirely unconvincing, given its lack of proper evidence.

[59] Budde 1872, p. 16.

Euclidean universe. The hypothesis of a Riemannian space, he said, "has usually been met with either an incredulous shake of the head or a contemptuous laughter," an attitude he found to be not entirely unwarranted. Indeed, Eugen Karl Dühring, a philosopher and amateur physicist, had the same year ridiculed non-Euclidean geometry as "mathematical mysticism" and attacked Gauss (this "son of a bricklayer") for having seriously proposed that the curvature of space might be detected by geodetic measurements.[60] Vaihinger's philosophical discussion of Zöllner's universe did not reveal much understanding of either thermodynamics or non-Euclidean geometry, nor of Zöllner's solution of Olbers' paradox.

Two years later, another philosopher dealt with Zöllner's hypothesis, this time more systematically and with greater insight. Wilhelm Wundt, a former assistant to Helmholtz, is primarily known as a pioneer of experimental psychology but his interests ranged widely, covering also many aspects of philosophy and the foundations of science. In his lengthy article of 1877, Wundt formulated the essence of the cosmological problem as he and many of his contemporaries conceived it: "Is the world finite or infinite in time? Is the world finite or infinite in space? Is the world finite or infinite in the mass of its matter?"[61] These possibilities he examined critically and systematically, using a substantial part of his paper to argue against Zöllner's model of a finite-space and finite-matter universe. One of his arguments was that this kind of universe would be subject to the heat death and consequently also have a beginning in time – contrary to Zöllner's assumption of an eternal universe. Apart from expressing general dissatisfaction with the non-intuitiveness of curved space, Wundt claimed that in such a space the geometrical and possibly also physical properties of bodies would change with their location in space, a claim that contradicted Helmholtz's principle of the "free mobility of rigid bodies."[62] Moreover, he argued for a fundamental space-time symmetry, in the sense that the weird concept of positively curved time followed as a consequence of Riemannian curved space. "Transcendental time is the necessary complement of transcendental space," he declared.[63]

---

[60] Dühring 1875, p. 67. He also attacked Riemann and Helmholtz, in no kinder language, for their belief in the "piquant nonsense" of transcendental geometry (Dühring 1877, p. 460; see also Cahan 1994). Dühring did not refer to Zöllner, whose view of a universe containing only a finite number of stars he shared, if for reasons entirely different from Zöllner's. Dühring's philosophical, political and scientific ideas were attacked by Friedrich Engels in his *Anti-Dühring* of 1878 and later in his posthumous *Dialektik der Natur*, and it is mostly through this connection he is known today (Kragh 2008, pp. 132-136). Incidentally, in the latter work, in a manuscript of 1878, Engels also attacked Zöllner for his advocacy of a spiritual fourth dimension. He did not mention the curved-space hypothesis.

[61] Wundt 1877, p. 81.

Wundt's own favourite universe was doubly infinite (in space and time), but with a finite amount of matter. Instead of having matter distributed uniformly, he argued that the density of cosmic matter decreased from the solar system: it would approach zero at infinity in such a way that the totality of matter in the universe was a finite quantity. In spite of denying the reality of Zöllner's curved universe, the philosopher-psychologist valued it from a philosophical point of view: "Zöllner's attempt to introduce the concept of transcendental space into the field of cosmology may possibly belong to the most fruitful ideas that had been for a long time proposed in cosmology."[64] He considered the attempt to be wrong, but a challenge to conventional wisdom that might open up a whole new way of thinking about physical space. And right he was. Wundt's paper in the *Vierteljahrschrift für wissenschaftliche Philosophie* was followed by an equally long discussion by Kurd Lasswitz, a German author and philosopher who at the time taught mathematics and physics at a gymnasium in Gotha.[65] Lasswitz claimed to have disproved Wundt's hypothesis of a finite mass in a universe of infinite space and time, but without adopting Zöllner's alternative. Although he discussed the Zöllner-Riemann curved space as a possible candidate for the structure of the universe, it was only to reject it.

---

[62] According to Helmholtz, the constant-curvature non-Euclidean geometries satisfied the criterion, necessary for any physical space, that motion do not imply change in form. He first examined which geometries can account for the invariance of bodies under transformation, and therefore our empirical measurements of objects, in a paper of 1868 (Helmholtz 1968, pp. 32-60). Although this was not the last word on the subject, Helmholtz's criterion was widely accepted until the end of the century.

[63] Wundt 1877, p. 115. The hypothesis of curved or cyclic time (rather than space-time) *is* weird. In such a world the space curvature does not vary periodically *in* time but time itself moves, as it were, on a circle. This kind of cyclic time has occasionally been discussed by philosophers, but it has not found any use in science and is generally thought to be absurd (Whitrow 1980, pp. 39-41).

Much of the German discussion about cosmology in the 1870s and 1880s was related, directly or indirectly, to the cultural struggle between "materialists" and Catholic thinkers that went on at the time. Generally speaking, while materialists and positivists were in favour of a spatially infinite universe, Catholics (and Christians generally) subscribed to a finite universe.[66] One might therefore believe that Christian scholars, many of them Jesuits, embraced the Zöllner-Riemann hypothesis, but this was not the case.

Consider Constantin Gutberlet, a prominent Catholic philosopher and theologian who in a number of publications carefully investigated the problem of a universe containing an infinite number of bodies. Based upon astronomical, mathematical and philosophical arguments, he concluded that such a universe was impossible. Thoroughly familiar with non-Euclidean geometry, including the works of Riemann and Zöllner, he nonetheless denied that physical space could be curved. The Euclidean nature of our local space was self-evident, he claimed, a fact that "no rational person can deny." And so it was with astronomical space: "As to the largest triangles that have been measured, and even to those which because of their size cannot be measured at all, no one can seriously claim that they might not be plane."[67] Consequently, Zöllner's idea of a closed space was deemed to be nothing but a mathematical abstraction. Gutberlet's dislike of Zöllner's hypothesis was not only rooted in its non-Euclidean feature, but even more so in Zöllner's adoption of infinite cosmic time. According to a Christian perspective, Zöllner's use of the principle of sufficient reason as an argument for eternity was invalid since it presupposes that the world is necessary rather than contingent.

---

[64]  Wundt 1877, p. 119.
[65]  Lasswitz 1877, followed by Wundt's reply on pp. 361-365. Zöllner did not respond to either Budde, Vaihinger, Wundt or Lasswitz.
[66]  The theme is detailed in Kragh 2008.

From a perspective very different from Gutberlet's, also the German mathematician Hermann Schubert dismissed Zöllner's belief that the visible world is contained in a four-dimensional space.[68] Schubert, who subscribed to positivist philosophy in the style of Mach, was unimpressed by Zöllner's astronomical use of the hypothesis. He seems not to have realized that its essence was a three-dimensional curved space and that it did not necessarily presuppose a hyperspace of higher dimension.

As a late example of the philosophical-cosmological discussion in German-speaking Europe, a work by the young Austrian physicist Arthur Erich Haas deserves mention. In a popular lecture given in Vienna in 1911, Haas discussed the traditional question of spatial and temporal infinity, concluding that the universe might well be finite in both respects. In this context, he referred to Zöllner's hypothesis, which at the time had fallen into oblivion. The hypothesis of a closed space had become more attractive, Haas suggested, because Olbers' paradox could no longer be explained away in terms of interstellar absorption. To explain the idea of a spherical space to his audience, mostly consisting of amateur astronomers, he used an analogy that had been standard since the 1870s, when Helmholtz introduced it: "Would it not be conceivable that a cosmic wanderer, on his way from the sun to Sirius, and then beyond it, finally returns after a very long journey to the sun? Based on our geometrical considerations, this would in no way be inconceivable."[69]

---

[67] Gutberlet 1908, p. 54. He first studied the philosophical consequences of non-Euclidean geometry in Gutberlet 1882.

[68] Schubert 1893, pp. 430-436. Schubert, who worked as a high school teacher in Hamburg, is best known for his contributions to enumerative geometry. In 1897 he published a popular book on mathematics which six years later was translated into English by Thomas McCormack as *Mathematical Essays and Recreations* (Chicago: Open Court Publishing Company, 1903). The book included on pp. 64-111 the 1893 article on four-dimensional space.

## 6. Some later developments, 1870-1900

The popularity of non-Euclidean geometry in the last decades of the nineteenth century manifested itself primarily in texts of a mathematical and philosophical nature. To the last category belonged John Stallo's *Concepts and Theories of Modern Physics*, an innovative and influential book in the new tradition of empirical positivism. The German-American author dealt competently and in critical detail with what he interchangeably called "metageometry" and transcendental geometry (or sometimes "pangeometry"), devoting a full chapter to Riemann's 1854 lecture and its consequences. He found that Riemann's celebrated essay was full of "crude and confused" sentences even in its original German, and that Clifford's clumsy translation – which was "no doubt made, not *by*, but *for*, Professor Clifford, by some one who had a very insufficient knowledge of German" – only increased the obscurity. When, in the following chapter, Stallo turned his attention to cosmology, he failed to combine the two subjects. Although familiar with Zöllner's *Natur der Cometen*, he had nothing to say about the hypothesis of a closed universe described by Riemann's geometry. The reason might have been that he denied that space could be an object of sensation, which in his view invalidated "the theory of the geometrical transcendentalists."[70]

---

[69] Haas 1912, p. 171.

Stallo's fellow-positivist, the famous Austrian physicist and philosopher Ernst Mach, seems to have had a greater appreciation of non-Euclidean geometry. Like Gauss and the other fathers of this kind of geometry, he did not believe in the a priori validity of the Euclidean system. On the other hand, since geometry always referred to objects of physical experience he found it highly improbable (but not impossible) that physical space could be anything but Euclidean. His various and somewhat ambiguous critical remarks about the new transcendental geometry were not directed at Lobachevsky, Bolyai and Riemann, but at the "grotesque fictions" which the many misunderstandings of their ideas had given rise to.[71] Mach dealt insightfully and at length with non-Euclidean geometry in *Space and Geometry* from 1906, a book based on essays previously published in *The Monist*. While expressing gratitude to the geometers "for the abolition of certain *conventional barriers* of thought," he doubted that their insight would be of any value to the physicist, who had no reason to abandon Euclidean geometry. He thought that space, as an object of experience, was as unlikely to satisfy the ideas of the geometers as matter was to satisfy "the atomistic fantasies of the physicist."[72]

---

[70] Stallo 1882, p. 232. On Stallo and the significance of his book, see Moyer 1983, pp. 3-32.
[71] Mach 1903, p. 27.
[72] Mach 1906, p. 136 and p. 141.

In his masterwork *The Science of Mechanics*, Mach paid tribute to the "very important" work of Riemann, "the significance [of which] can scarcely be overrated." At the same time he dismissed all ideas of a fourth dimension as nothing but an "opportune discovery for the spiritualists and for theologians who were in a quandary about the location of hell."[73] Mach was as adamant in his rejection of geometries with more dimensions than three as he was silent about the geometrical structure of astronomical space. The German-American author, publisher and philosopher Paul Carus was a friend of Mach and a leading figure in the positivist movement. The editor of the journal *The Monist*, he was well acquainted with the contemporary discussions of space, yet he seems to have completely misunderstood the basic ideas of non-Euclidean geometry. In a book of 1908 he wondered what lied outside Riemann's boundless but finite spherical space: "Would the name 'province of the extra-spatial' perhaps be an appropriate term? I do not know how we can rid ourselves of this enormous portion of unutilized outside room."[74]

What about the astronomers, then? I am aware of only two (or, including Peirce, two-and-a-half) professional astronomers who in the three decades after 1870 expressed concern with non-Euclidean space, and none of them went much beyond uncommitted comments. The Irish astronomer Robert Stawell Ball was in 1874 appointed Royal Astronomer of Ireland and professor at the University of Dublin, which included the directorship of the Dunsink Observatory. From 1892 to his death in 1913 he served as Lowndean Professor of Astronomy and Geometry at Cambridge University.[75] Much of his scientific work was devoted to mathematical physics (in particular the "theory of screws"), but while at Dunsink he also directed a large-scale observational research programme in determining stellar parallaxes. Among the problems that faced astronomers in this area was the choice of comparison stars for parallax measurements by taking into account the proper motions of the stars. Ball and his collaborators paid particular attention to the star 61 Cygni that Bessel had originally used in his discovery of the annual parallax. In a lecture given to the Royal Institution on 11 February 1881, he discussed the complex questions of comparison stars and proper motions in relation to parallax measurements. Then, at the end of the lecture, he alluded to "the nature of space" in the following brief way:

> If space be hyperbolic the observed parallax is smaller than the true parallax, while the converse must be the case if space be elliptic. The largest triangle accessible to our measurements has for base a diameter of the earth's orbit, and for vertex a star. If the *defect* of the sum of the three angles of such a triangle from two right angles be in any case a measurable quantity, it would seem that it can only be elicited by observations of the same kind as those which are made use of in parallax investigations.[76]

Ball was well acquainted with non-Euclidean geometry, but his remarks in the 1881 address had the character of an afterthought rather than a serious proposal for investigating the geometry of space by astronomical means. He did not elaborate on the subject in his later work. In his best-selling *The Structure of the*

---

[75]  On Ball's life and career, see Ball 1915 and the obituary notice in *Proceedings of the Royal Society A 91* (1915), xvii-xxii.

[76]  Ball 1881.  What Ball called the "true parallax" is the angle under which the radius of the orbiting earth appears for an observer located at a star; the "observed parallax," on the other hand, is half the annual change of the angular distance between the star and some comparison star close to it.

*Heavens*, a popular book of 1886, he dealt at length with methods of parallax determinations, but without mentioning the possibility of curved space.[77] The same year he wrote a paper on measurement distances in elliptic space, which might have been an opportunity to draw a connection to astronomy. But the paper was purely mathematical and without any indication that the subject might be of relevance in astronomical contexts.[78]

Characteristically, Ball's guarded preference for a closed cosmic space turned up in his popular publications only. In *The High Heavens* of 1893, he discussed whether space is finite or infinite, a question which "is rather of a metaphysical complexion" and "depends more on the facts of consciousness than upon those of astronomical observation."[79] Having argued that the number of matter particles in the universe must be finite, he proceeded to space itself and the possibility of "a space which is finite in dimensions" – spherical space, that is. Although Ball did not explicitly support a positively curved space, he stressed that it was consistent and intuitively acceptable. Indeed, he expressed sympathy with the hypothesis, which "provides the needed loophole for escape from illogicalities and contradictions into which our attempted conceptions of [infinite] space otherwise land us."[80]

In this context may be mentioned also the American mathematician James Edward Oliver, professor at Cornell University, who according to George Halsted was "a pronounced believer in the non-Euclidean geometry." Halsted recalled how Oliver tried to convince him that astronomical evidence pointed to space being closed. At one occasion Oliver "explained a plan for combining stellar spectroscopy with ordinary parallax determinations, and expressed his disbelief that C. S. Pierce [sic] had proved our space to be of Lobachévsky's kind, and his conviction that our universal space is really finite, therein agreeing with Sir Robert Ball."[81] Unfortunately, it remains unknown what Oliver's ideas were about, more precisely.

---

[77] Ball 1886a, pp. 441-460.
[78] Ball 1886b. As we shall see, the disinclination to connect astronomy and non-Euclidean geometry in more than a rhetorical way was a feature also in Newcomb.
[79] Ball 1893, p. 247
[80] Ibid., p. 252. Carus 1908, p. 123, referred to Ball as an advocate of Riemannian space, a concept he found to be strange and conceptually contradictory.

Like Ball, the distinguished American astronomer Simon Newcomb combined mastery of the practical and mathematical sides of astronomy. In 1877, at a time when he had just become superintendent of the Nautical Almanac Office, he published a paper in Crelles' *Journal* on the geometry of space with positive curvature. Except by pointing out that "there is nothing within our experience which will justify a denial of the possibility that the space in which we find ourselves may be curved in the manner here described,"[82] he did not relate his investigation to physics or astronomy. Newcomb's space was not quite the same as Riemann's, but described by what Felix Klein in his memoir of 1871 on non-Euclidean geometry called "elliptic geometry." While in spherical or Riemannian space all geodesics from a given point intersect again at a distance $\pi R$, in elliptic space two geodesics can have only one point in common. In the latter case the largest possible distance between two points is $\frac{1}{2}\pi R$, whereas it is $\pi R$ in the spherical case. Both spaces are finite, but for the same radius of curvature the volumes differ. The volume of a spherical space is $2\pi^2 R^3$, and it is $\pi^2 R^3$ for the elliptic case.

Although not considered to be of astronomical relevance at the time, Newcomb's paper was of importance to the development of mathematics: it inspired 31-year old Wilhelm Killing, at the time professor at a seminary college in Braunsberg, to take up non-Euclidean geometry, a field that became a lifelong research interest to him. Convinced of the mathematical validity of Newcomb's version of closed space, in 1878 Killing proved that there are three distinct types of constant-curvature non-Euclidean geometry: Lobachevsky's hyperbolic space, Riemann's spherical space, and the elliptic space first considered by Newcomb.[83]

---

[81] Halsted 1895, p. 545. Oliver seems not to have published his plan nor, indeed, anything about non-Euclidean geometry.
[82] Newcomb 1877, p. 299.

Newcomb mentioned at various occasions the possibility of curved physical space, but in popular contexts only and without taking it too seriously. He seems to have been reluctant to part with the infinite Euclidean space. In the widely read *Popular Astronomy*, a book of 1878 that over the next twenty years was published in many editions, he discussed what would happen with the heat of the sun. Would it forever be lost? Or would it, if space were curved, return to the sun? In agreement with what he had said the year before, he wrote: "Although this idea of the finitude of space transcends our fundamental conceptions, it does not contradict them and the most that experience can tell us in the matter is that, though space be finite, the whole extent of the visible universe can be but a very small fraction of the sum total of space." But Newcomb did not believe in the possibility of a positively curved space in which the solar heat would return to its source. On the contrary, he dismissed the hypothesis as "too purely speculative to admit of discussion."[84]

Many years later, in an address to the American Mathematical Society, Newcomb dealt in a general way with what he called the philosophy of hyperspace, with which he meant any geometrical system transcending the ordinary Euclidean conception of space. Thus, "hyperspace" included extra-dimensional as well as non-Euclidean spaces, two concepts that otherwise had nothing in common: "Curved space and space of four or more dimensions are completely distinct in their characteristics, and must, therefore, be treated separately."[85] As Newcomb pointed out, the hypothesis of curved space was testable, if more in principle than in practice: "Unfortunately, we cannot triangulate from star to star; our limits are the two extremes of the earth's orbit. All we can say is that, within those narrow limits, the measures of stellar parallax give no indication that the sum of the angles of a triangle in stellar space differs from two right angles." He continued with an argument that ruled out elliptic space as more than a speculation, as seen from the astronomer's perspective:

> If our space is elliptical, them, for every point in it – the position of our sun, for example – there would be, in every direction, an opposite or polar point whose locus is a surface at the greatest possible distance from us. A star in this point would seem to have no parallax. Measures of stellar parallax, photometric determinations and other considerations show conclusively that if there is any such surface it lies far beyond the bounds of our stellar system.[86]

---

[83] Killing 1878, which was a purely mathematical paper.
[84] Newcomb 1878, pp. 504-505. On Newcomb's philosophy of science, a mixture of positivism and operationalism, see Moyer 1983, pp. 83-96.

Newcomb's cautious ideas about non-Euclidean space form an instructive contrast to those of his compatriot and friend, Charles Sanders Peirce. Although today mostly known as a philosopher and founder of semiotic theory, as a young man Peirce, a polymath if there ever were one, was primarily recognized as a promising astronomer and chemist. While at Harvard College Observatory he did important work in photometry and spectroscopy, and he was among the first to study the spectrum of an aurora. Elected a member of the U.S. National Academy of Science in 1877, he spent most of his professional career as a practicing scientists associated with the United States Coast and Geodetic Survey.

---

[85] Newcomb 1898, p. 2. Presidential address to the American Mathematical Society of 29 December 1897. Although not a believer in the fourth dimension in the style of Zöllner, neither did Newcomb rule it out as absurd. He speculated that the ether might provide a bridge between our world and the fourth dimension, "For there is no proof that the molecule may not vibrate in a fourth dimension." Moreover, "perhaps the phenomena of radiation and electricity may yet be explained by vibration in a fourth dimension." See also Newcomb 1906, which included an essay on "The Fairyland of Geometry" (pp. 155-164) covering much of the same ground as his 1897 address, but without discussing astronomical and physical aspects of hypergeometry.

[86] Newcomb 1898, p. 7.

Contrary to the four years older Newcomb, Peirce was convinced that our space is non-Euclidean – indeed *must* be non-Euclidean – a claim he supported with both philosophical and observational arguments.[87] In letters and manuscripts from the years 1891-1902 Peirce investigated various aspects of the structure of space, which led him to conclude that it was either of the Lobachevskian or the Riemannian kind. "Only the exactest measurements upon the stars can decide," he wrote in a manuscript from about 1902.[88] In an earlier paper he discussed the question in terms of stellar parallaxes, although without suggesting an answer to the sign of space curvature: "I think we may feel confident that the parallax of the furthest star lies somewhere between – 0.''05 and + 0.''15, and within another century our grandchildren will surely know whether the three angles of a triangle are greater or less than 180°, – that they are *exactly* that amount is what nobody ever can be justified in concluding."[89] However, Peirce seems to have had a predilection for hyperbolic space, such as is evidenced from his manuscripts and correspondence with Newcomb in the early 1890s.

Thus, in one of his manuscripts of 1891 he listed no less than fifteen "methods of investigating the constant of space" that included parallax measurements, ideas of stellar evolution, the proper motions of stars, and Doppler shifts in stellar spectra. In addition, "the relative numbers of stars of different magnitudes depend on the constant of space."[90] In another manuscript: "I made the necessary computations for a selection of stars. The result was markedly in favor of the hyperbolic geometry."[91] This is also what he reported in a lengthy letter to Newcomb, where he convinced himself – and in vain tried to convince Newcomb – that astronomical data provided support for his "attempt to make out a negative curvature of space." Although realizing the hypothetical nature of his conclusion, he had no doubt of its significance:

> The discovery that space has a curvature would be more than a striking one; it would be epoch-making. It would do more than anything to break up the belief in the immutable character of mechanical law, and would thus lead to a conception of the universe in which mechanical law should not be the head and centre of the whole. It would contribute to the improving respect paid to American science, were this made out here. … In my mind, this is part of a general theory of the universe, of which I have traced many consequences, – some true and others undiscovered, – and of which many more can be deduced; and with one striking success, I trust there would be little difficulty in getting other deductions tested. It is certain that the theory if true is of great moment. [92]

---

[87] For a critical analysis of these arguments, see Dipert 1977.
[88] Dipert 1977, p. 407.
[89] Peirce 1891, p. 174.

To investigate the matter more closely, Peirce estimated he needed a research grant corresponding to six to nine months of work. "The question is, can some appropriation be made, or some millionaire be found, to pay $3000 for this,

---

[90] Manuscript of 24 March 1891, in Peirce 2010, pp. 229-230.

[91] Manuscript of 1894, quoted in Dipert 1977, pp. 411-412. In 1892 Peirce reviewed George Halsted's translation of Lobachevski's *Geometrical Researches on the Theory of Parallels* (Austin: University of Texas, 1891). The review, which appeared in *The Nation* 54 (11 February 1892), 116, is reproduced in Peirce 2010, pp. 271-274.

[92] Undated letter of 1891, reproduced in Eisele 1957, pp. 421-422. The quotation gives an indication of Peirce's general and wide-ranging philosophical system, as published in a series of five papers in *The Monist* 1891-1893 and of which Peirce 1891 was the first. The papers are reproduced in, for example, Peirce 1972, pp. 159-260. According to Peirce's system, even the fundamental laws of nature were probabilistic and inexact, and they might even evolve in time.

$2000 for six months of my work and $1000 for an assistant?" Peirce's optimism was short-lived, such as indicated in a later letter to Newcomb: "I have for the present given up the idea that anything can be concluded with considerable probability concerning the curvature of space."[93] Newcomb welcomed Peirce's more agnostic attitude, which he took to be support of his own view, namely, "that all philosophical and logical discussion is useless."[94] This was definitely *not* a view shared by Peirce!

Among the few attempts in the nineteenth century to conceive celestial space as non-Euclidean, Peirce's was the most elaborate and serious one. However, it had no influence at all, primarily because he did not publish his arguments in journals read by most astronomers and mathematicians. Although they were known to some American scientists, they failed to convince them. As Newcomb wrote him, "the task of getting the scientific world to accept any proof that space is not homoloidal, is hopeless, and you could have no other satisfaction than that of doing a work for posterity." With regard to the possibility of a research grant, his reply was equally disappointing: "It is, I believe, unusual if not unprecedented, to pay an investigator to do work of his own out of trust funds for the advancement of science, at least among us. I do not know where to look for funds to do this with." [95]

● The parallax methods discussed by Lobachevsky, Ball, Peirce and other scientists to estimate the curvature of space were variations of the same approach, which in its essence can be summarized as follows. Consider a triangle ABC, where a star is located at A, and B and C are the positions of the earth half a year apart in its orbit around the sun. The base BC is one astronomical unit, and the angles β and γ in the triangle (at corners B and C, respectively) are known observationally. In the Euclidean case the parallax is the angle $p = \pi - (\beta + \gamma) = \alpha$. For any space of constant curvature, Euclidean or not, it can be shown that the difference of the angle sum from 180° for a triangle is given by the product of the curvature and the area of the triangle:

$$\alpha + \beta + \gamma - \pi = K\delta,$$

where δ is the area of the triangle. The quantity $K$ is the curvature, related to the radius of curvature $R$ by

$$R^2 = \frac{k}{K}, \;\; k = \pm 1,$$

where $k = +1$ refers to spherical space and $k = -1$ to hyperbolic space. It follows that the parallax in curved space can be expressed as

$$p = \pi - (\beta + \gamma) = \alpha - K\delta$$

For the hyperbolic case the parallax of distant stars ($\alpha \cong 0$) will thus remain positive. In other words, if stars are observed with a zero parallax, meaning a parallax of the same magnitude as the error of observation, the error will give an upper limit to the numerical curvature. In the case of a distant star in spherical space ($K > 0$), the sum β + γ will be greater than π and so the parallax should be negative. If no stars are observed with $p < 0$, the error of observation will again give an upper limit to $K$.

There are other ways of illustrating the difficulty of measuring the curvature of space. In spherical space the length of a circle of radius $r$ is

$$l = 2\pi R \sin\frac{r}{R},$$

or, for $r \ll R$,

$$l \cong 2\pi r \left[ 1 - \frac{1}{6}\left(\frac{r}{R}\right)^2 \right]$$

Similarly for the area $A$ of a spherical cap:

$$A(r) = 2\pi R^2 \left( 1 - \cos\frac{r}{R} \right) \cong \pi r^2 \left[ 1 - \frac{1}{12}\left(\frac{r}{R}\right)^2 \right]$$

Thus, measurements of $l$ and $r$, or $A$ and $r$, permit the radius of curvature $R$ to be determined. The problem, recognized by nineteenth-century astronomers and mathematicians, is that it requires enormous distances. For example, for the very large distance $r = 0.1R$ the length of a cosmic circle in spherical space differs from the Euclidean case by only 0.2%, and for the area the difference is 0.1%.  ●

## 7. The pre-relativistic cosmos

Around the turn of the century discussions of non-Euclidean geometry were widespread. Although predominantly a business for mathematicians and philosophers, they can also be found elsewhere, for example among physicists and astronomers. As one more example, consider the famous Austrian physicist Ludwig Boltzmann, who dealt with the subject in his lectures on theoretical physics and also in the lecture course on natural philosophy he gave in Vienna in the last years of his life. For example, he referred to "the suggestion that the distances of fixed stars may perhaps be constructed only in a non-Euclidean space of very small curvature." Moreover, he connected the curvature with the law of inertia, namely, "a moving body not acted on by forces would then after aeons have to return to its previous position if the curvature is positive."[96]

Boltzmann elaborated on the subject in his lectures on natural philosophy, where he rejected Kant's a priori conception of space and argued that curved space might well be consistent with experience. In his advocacy of the possibility of curved space he referred to heavenly triangles with stars in their vertices, but without going further than Lobachevski and some of his followers. "The spherical non-Euclidean space is completely closed in itself; it is not infinite, but has some finite size," he said. "If we know how large the triangles must be to correspond to a certain deviation from the sum of angles 180°, then we could also construct the size of the entire universe [*ganzen Raumes*]. We would then have a space which ends nowhere and as a whole returns into itself."[97] He thought this was a perspective that offered "enormous logical advantages." But one thing is logic, another is empirical reality. While in one of his lecture notes Boltzmann considered a closed universe to be not only possible, but also probable, in a later note he held it to be "not likely, yet it is a possibility that measurements of the stars will prove space to be non-Euclidean."[98] This was a view common at the time, and one which can be found in the scientific as well as popular literature.

---

[96]  Boltzmann, lectures on the principles of mechanics, as quoted in Cercignani 1998, p. 167.

To the extent Boltzmann has a name in the history of cosmology it is not due to his vague ideas of curved space but because of his speculations about the cosmological significance of the second law of thermodynamics. As he pointed out in a paper of 1895, in a sufficiently large universe there will be "worlds" in which the entropy decreases rather than increase, a result of the statistical nature of the entropy law.[99] He did not, at this time, comment on the geometrical structure of the universe; and when he, a few years later, considered the possibility of a non-Euclidean cosmic space, he did not connect it with his earlier entropic speculations.

While the possibility of space being non-Euclidean does not seem to have aroused interest among French astronomers, their colleagues in mathematics did occasionally consider the question, if in an abstract way only. As mentioned, many scientists were of the opinion that the geometry of space could be determined empirically, at least in principle. However, not all agreed, and especially not in France. On the basis of his conventionalist conception of science, Henri Poincaré argued that observations were of no value when it came to a determination of the structure of space. He first published his idea of physical geometry being a matter of convention in a paper of 1891 entitled "Les géométries non-euclidiennes," and later elaborated it on several occasions.[100]

---

[97] Fasol-Boltzmann 1990, p. 215. See also Tanaka 1999 for Boltzmann's interest in non-Euclidean geometry and other parts of abstract mathematics.
[98] Fasol-Boltzmann 1990, p. 215 and p. 255.
[99] Boltzmann 1895.

According to Poincaré, observations can only tell us about the relations that hold among material objects such as rigid rods and the properties of rays of light. For example, if the sum of angles in a celestial triangle was found by astronomical measurements to be 200°, one might assume the physics of light propagation to be correct and change to a spherical geometry; but one might also choose to maintain Euclidean geometry by changing the theory of how light propagates through space. This line of reasoning may first have been proposed by Lotze in his *Metaphysik* of 1879, as part of his argument that space cannot show properties that contradict our intuitions.[101] As far as Poincaré was concerned, the geometry of space was not something that could be determined objectively: "One geometry cannot be more true than another; it can only be more convenient."[102] Because Poincaré found Euclidean geometry to be the most simple and convenient system, he saw no reason to consider other candidates for the structure of space.

Auguste Calinon, a French mathematician and philosopher, was a correspondent of Poincaré and shared his interest in the foundations of mechanics and geometry. Although influenced by Poincaré's conventionalism, he did not believe that Euclidean geometry was always to be preferred because of its simplicity. Generally, very few geometers, whether in France or elsewhere, adopted the extreme conventionalist view of Poincaré.[103] As to Calinon, he stressed that different geometries are not simply equivalent and that it is legitimate to ask about the particular geometry that is realized in the physical world. In a paper of 1889 he pointed out that, "In order to know which of these [different] spaces contains the bodies we see around us, we must necessarily look to experience."[104] He distinguished between three hypotheses concerning the geometry of the universe: (1) it is strictly Euclidean; (2) it is slightly non-Euclidean, but with constant curvature; (3) the curvature of space varies in time. The latter and more general hypothesis he spelled out as follows: "Our space realizes different geometric spaces successively in time; in other words, our spatial parameter [space constant] varies with time, either by deviating more or less from the Euclidean parameter, or by oscillating around a given parameter very near the Euclidean parameter."[105]

---

[100]  It was first translated into English as Poincaré 1892.

[101]  See Russell 1897, p. 100, and Torretti 1978, p. 289.

[102]  See the essay on "Non-Euclidean Geometries," pp. 35-50 in Poincaré 1952a, first published in 1902. In another essay Poincaré illustrated his conventionalism by imagining a uniform expansion of the universe, concluding that this "immense convulsion" would have no observable effects whatever; the reason being that the dimensions of everything, measuring instruments included, would increase in the same ratio. Poincaré 1952b, p. 94. Although not an advocate of conventionalism, in his extensive article on "Geometry" in the 1911 edition of *Encyclopedia Britannica* (co-authored with Alfred N. Whitehead), Bertrand Russell came to the same conclusion with regard to astronomical evidence for curved space. Online edition: http://www.1911encyclopedia.org/Geometry.

Although Clinon spoke of astronomical measurements of celestial triangles as a "mode of verification" of Euclidean geometry, he may not have believed that a non-Euclidean structure of space might ever be revealed observationally. "All that can legitimately be concluded," he said, "is that the differences which might exist between Euclidean geometry and that realized by the universe are due to experimental error."[106]

In another paper Calinon argued in favour of an unavoidable geometric indeterminacy of the universe, a result he ascribed to our measurements being always inexact and limited to a small part of the universe as a whole. The indeterminacy was thus of an epistemological rather than ontological nature. Astronomical and other empirical problems might be approached with the kind of geometry most suited to produce a simple solution, he said, and the choice of geometry might vary from one problem to another, and from one area of the universe to another. Any geometry might be applied to the actual world by a suitable hypothesis concerning the course of rays of light. "By itself, space is neither finite nor infinite," he declared. "What is finite or infinite is solely this or that particular representation that determines the geometry of our universe."[107] Calinon applied the same kind of reasoning to the laws of nature, arguing that they would differ in accordance with the chosen geometrical representation. For example, one might assume that space differed from the Euclidean form at very large distances – say beyond Sirius – and that Newton's law of gravitation might consequently need to be modified. "We may therefore very well conceive that at such large distances the law of attraction could find its simplest expression in another geometric representation of the universe, different from the Euclidean representation."[108] This was a view in agreement with Poincaré's version of conventionalism.

---

[103]  This is convincingly shown in Scott 1997.
[104]  Calinon 1889, p. 594, which was followed by Calinon 1891. An English translation of the first paper is published in Čapek 1976, pp. 297-303. See also Torretti 1978, pp. 272-276.
[105]  Calinon 1889, p. 595.
[106]  Ibid.

Calinon's arguments were of a philosophical nature that failed to appeal to either astronomers or physicists. They did however attract the attention of Bertrand Russell, who discussed them in his doctoral dissertation of 1897 on the foundations of geometry. The English mathematician-philosopher dismissed Calinon's hypothesis of a space constant varying in time with the argument that "this would involve a causal connection between space and other things, which seems hardly conceivable."[109] He also objected that astronomical measurements would not be trustworthy if space changed during the time it takes a signal to reach the earth from some distant star. If the curvature of space might conceivably change in time, why not in cosmic space as well? The idea was occasionally suggested, but, to my knowledge, in popular contexts only. Calinon mentioned it, and he was not the first. While Riemann and Clifford had entertained the idea of a varying space curvature, they had restricted it to microphysics. According to Tait, professor of natural philosophy in Edinburgh, the investigations of "mathematicians of the highest order," such as Riemann and Helmholtz, implied the more drastic notion of space varying over astronomical distances. In a popular book of 1876 he speculated:

> The result of their inquiries leaves it as yet undecided whether space may or may not have precisely the same properties throughout the universe. … So it is possible that, in the rapid march of the solar system through space, we may be gradually passing to regions in which space has not precisely the same properties as we find here – where it may have something in three dimensions analogous to curvature in two dimensions – something, in fact, which will necessarily imply a fourth-dimension change of form in portions of matter in order that they may adapt themselves to their new locality.[110]

---

[107] Calinon 1893, p. 605.
[108] Ibid., p. 607.

Somewhat similar speculations, but more general and not relating to astronomical space, appeared in other popular works. Clifford's posthumous *Common Sense of the Exact Sciences* was completed by Karl Pearson, who was responsible for the chapter on curved spaces in which different kinds of variation of the curvature of space was discussed. According to Pearson, "The hypotheses that space is not homaloidal, and again, that its geometrical character may change with time, may or may not be destined to play a great part in the physics of the future; yet we cannot refuse to consider them as possible explanations of physical phenomena."[111] He was silent about astronomical phenomena.

---

[109] Russell 1897, pp. 112-113. Of course, the kind of causal connection between "space and other things" that Russell found inconceivable was soon justified by Einstein's general theory of relativity.
[110] Tait 1876, p. 5.

A contemporary of Calinon, the mathematician Paul Barbarin, professor at the Lyceum of Bordeaux, was a prolific writer on non-Euclidean geometry.[112] Contrary to Poincaré, he was an empiricist in the sense that he believed that the geometry of space was a question that could, and could only, be determined observationally. This is what he argued in a widely acclaimed book of 1902, *La géometrie non-euclidienne*, which included a chapter on "géométrie physique."[113] According to the French geometer, measurements of very small stellar parallaxes indicated that the radius of curvature exceeded 400,000 AU, which made him conclude that our part of the universe might possibly be curved. On the other hand, it might just as well be Euclidean, and from a practical point of view there was not as yet any means of distinguishing between the two possibilities. Barbarin optimistically expressed his belief that the problem would be solved in the future, thanks to the rapid progress in observation technology.

A paper of 1900 on "The Parameter of the Universe" (meaning the constant radius of space curvature) illustrates his awareness of the problem of the geometry of real space, but also the limitations of his mathematical perspective. Whereas, from the point of view of the mathematician, no geometry was truer than another, Barbarin noted that the physicist or astronomer did not have the same kind of luxurious freedom. Concerned with the space of the actual universe, they had to confront the question: "In the parts of the universe directly accessible to measurements, which geometrical system appears to be realized or at least to be best approximated?"[114] Barbarin approached the question by deriving formulae for celestial triangles that could in principle distinguish between the three geometries associated with the names of Euclid, Lobachevsky and Riemann. However, he had to admit that his formulae were of no practical value as they relied on angle measurements much more precise than 0″.01.

---

[111]  Clifford 1885, p. 226. According to Pearson's foreword, he was responsible for pp. 116-226. Clifford died in 1879, only 34 years old.
[112]  On Barbarin, see Halsted 1908.
[113]  Barbarin 1902, pp. 81-86.

As illustrated by its complete lack of astronomical data, Barbarin's paper was a geometrical exercise rather than a contribution to astronomy. It did nothing to change the general opinion at the time, such as summarized by Russell:

> Though a small space-constant is regarded as empirically possible, it is not usually regarded as probable; and the finite space-constants with which Metageometry is equally conversant, are not usually thought even possible, as explanations of empirical fact.[115]

Barbarin's paper may be contrasted to another paper of 1900, written by the 26-year-old German astronomer Karl Schwarzschild, a paper which has rightly been called "a high-water mark of practical non-Euclidean geometry as linked with classical astronomical physics."[116]

Schwarzschild was a student of the prominent astronomer Hugo von Seeliger, who had studied astronomy at the University of Leipzig under Zöllner and Karl Bruhns and since 1882 worked as a professor in Munich. A versatile astronomer who was equally at home in the mathematical, observational and philosophical aspects of his science, Seeliger was much interested in cosmological problems. In 1895 he proved that an infinite Euclidean universe with a roughly uniform mass distribution cannot be brought into agreement with Newton's law of gravitation, a problem known as the gravitation paradox, and in later works he studied the cosmic entropy problem and Olbers' paradox. Of interest in the present context, Seeliger denied that non-Euclidean geometry could be of any use in understanding physical, astronomical and cosmological questions. According to him, space in itself had no properties at all. In agreement with Poincaré, he warned against "the common and therefore very fatal misapprehension that one … [is] able to decide by measurement which geometry is the 'true' one, or even, which space is the one we live in."[117]

---

[114] Barbarin 1900, p. 71.
[115] Russell 1897, p. 53.
[116] North 1990, p. 78. The leading American cosmologist Howard Percy Robertson, famous for the Robertson-Walker metric of general relativity, called Schwarzschild's attempt to evaluate the curvature of space "so inspiring in its conception and so beautiful in its expression" (Robertson 1949, p. 321).

Young Schwarzschild did not agree with the view of his distinguished teacher, such as he made clear in an article based on a lecture given on 9 August 1900 to the *Astronomische Gesellschaft* in Heidelberg.[118] His aim was the same as that of Gauss and Lobatchevsky long before him, to determine the structure of space from observations – something which Poincaré and Seeliger deemed impossible in principle.

---

[117] Seeliger 1913, p. 200. In a paper of 1909, Seeliger questioned the validity of scientific laws to the universe as a whole, not only the law of gravitation but also the second law of thermodynamics. What he did not question, was the "law" that space is Euclidean (Seeliger 1909). For Seeliger as a cosmologist, see Kragh 2007, pp. 108-113.

[118] The following year, 1901, Schwarzschild was appointed professor of astronomy and director of the observatory at Göttingen University – the same position Gauss had occupied – and in 1909 he became director of the Astrophysical Institute at Potsdam near Berlin. Apart from his seminal contributions to astronomy and astrophysics, he also did very important work in electrodynamics (electron theory), quantum theory (Stark effect, simple molecules) and general relativity theory (Schwarzschild solution of spherical space-time, later accommodated in the physics of black holes). He died in 1916, while serving at the Russian front. See Dieke 1975.

Besides the Euclidean space, Schwarzschild discussed the hyperbolic space, having a constant negative curvature, and the closed elliptical space. Although the spherical space was better known to his audience, he argued that it, contrary to elliptical space, would lead to physically unacceptable consequences. As possible observational tests he proposed not only the classical parallax measurements but also star counts relating the number of stars to their brightness. (He was unaware that Peirce had made the same proposal a decade earlier.) While in Euclidean space the parallax $p$ of a star infinitely far away is zero, in hyperbolic space there will be a minimal non-zero parallax that decreases with the (imaginary) curvature radius $R$. Let the radius of the orbit of the earth be $r$, then $p \geq r/R$, as shown already by Lobachevsky in 1829. Thus, a measurement of the smallest known parallax imposes a lower limit on $R$. Schwarzschild estimated $p_{min} \cong 0''.005$ from which he concluded that $R > 4 \times 10^6$ AU. The bound, corresponding to about 20 parsec or 60 light years, was an order of magnitude higher than the one estimated by Lobachevsky. Schwarzschild commented: "Thus the curvature of the hyperbolic space is so insignificant that it cannot be observed by measurements in the planetary system, and because hyperbolic space is infinite, like Euclidean space, no unusual appearances will be observed on looking at the system of fixed stars."[119]

In the case of elliptical space there are no infinite distances, and every parallax, including $p = 0$, corresponds to a finite distance. The relevant formula replacing $p \geq r/R$ is

$$\cot \frac{d}{R} = p \frac{R}{r} \, ,$$

---

[119] Schwarzschild 1900, p. 342, with a rather poor English translation in Schwarzschild 1998. For the paper, and Schwarzschild's approach to cosmology, both classical and relativistic, see Schemmel 2005.

where $R$ is real and $d$ is the distance from the object (star) to the observer along a geodesic. Contrary to the hyperbolic case, "it is a mistake to believe that a limit for $R$ can be found simply from measurements of the parallax of fixed stars." Therefore, physical considerations are needed to determine the minimal value of $R$. The idea that metrical arguments are insufficient to determine space curvature, and that they have to be replaced or supplemented with physical arguments, had previously been suggested by Newcomb and Peirce, apparently unknown to Schwarzschild. Based upon star catalogues he argued that all stars having a parallax smaller than 0".1 were located within a finite volume, and from this, and by assuming a uniform distribution of the stars, he concluded that $R \cong 1.6 \times 10^8$ AU $\cong 10^4$ light years.

Although Schwarzschild's closed and finite world provided a logical solution to Olbers' paradox, and was in this respect similar to Zöllner's world, Schwarzschild did not refer to the famous paradox. Yet it entered indirectly, by the infinite trips that stellar light rays would make around the small universe. Schwarzschild knew, as already Helmholtz had pointed out in 1870, that in elliptic space a ray of light will return to its starting point after having traversed the world. We should therefore expect to see an antipodal image of the sun, a "counter-sun" identical to our ordinary image of it but in the opposite direction. Of course, no such second image of the sun is observed, a problem that Schwarzschild solved, or explained away, by assuming a suitable absorption of light in interstellar space. He summed up his result as follows:

> One may, without coming into contradiction with experience, conceive the world to be contained in a hyperbolic (pseudo-spherical) space with a radius of curvature greater than 4 000 000 earth radii, or in a finite elliptic space with a radius of curvature greater than 100 000 000 earth radii, where, in the last case, one assumes an absorption of light circumnavigating the world corresponding to 40 magnitudes.

Schwarzschild saw no way to go further than this rather indefinite conclusion and decide observationally whether space really has a negative or positive curvature, or whether it really is finite or infinite. Nonetheless, from a philosophical point of view he preferred a closed universe. It would, he said, be "satisfying to reason" if we could conceive of

> … space itself as being closed and finite, and filled, more or less completely, by this stellar system. If this were the case, then a time will come when space will have been investigated like the surface of the earth, where macroscopic investigations are complete and only the microscopic ones need continue. A major part of the interest for me in the hypothesis of an elliptic space derives from this far reaching view.[120]

In his spirited discussion of a curved cosmic space there was one assumption that he, contrary to Zöllner nearly thirty years earlier, failed to mention, namely, that the universe had existed in an eternity of time. But this was an assumption rarely questioned or even mentioned at the time, and one that also went unquestioned in the early relativistic models of the universe. It was also tacitly assumed by Paul Harzer, professor of astronomy at the University of Kiel, in an interesting lecture of 1908 that in some respects went further than Schwarzschild's analysis.

Like Schwarzschild, Harzer questioned the conventional wisdom that the finite stellar system was located in an infinite Euclidean space. He emphasized that only if this kind of space were assumed would there be a contradiction between an unbounded and finite universe. "Only experience can tell us which value of the geometrical curvature is valid for the actually existing universe," he said, presenting geometry as an empirical science on par with analytic mechanics.[121] This was little more than a paraphrase of what Gauss had said nearly a century ago, but Harzer proceeded to discuss some of the astronomical consequences of a closed space, both in its spherical and elliptical versions.

---

[120]  Scharzschild 1900, p. 342. Schwarzschild's attitude was strikingly similar to the one later expressed by Georges Lemaître in the context of relativistic cosmology. Lemaître was epistemically committed to finitude, arguing that only in this case would the universe in its totality be comprehensible to the human mind (Kragh 2008, p. 195). While Lemaître's finitism was to some extent rooted in his religious faith, this was not the case with Schwarzschild.

In his paper of 1900, Schwarzschild had stated that "the theory could be developed for curved space on the basis of the same general principles that professor Seeliger has used for Euclidean space, and comparison with experience could possibly show that the assumption of a space with a particular curvature leads to the simplest model of the distribution of the stars."[122] This is what Harzer did, although without arriving at a definite conclusion. While Seeliger had based his "statistical cosmology" on the assumption of a Euclidean space, Harzer transformed his calculations to a space of constant positive curvature.[123] In this way he arrived at a modified picture of the stellar Milky Way universe, now enclosed in a finite cosmic space of a volume about 17 times that of the stellar system. As to this stellar system, it contained the same number of stars but was compressed to a size roughly one half of that it had in the case of an infinite Euclidean space. The size of the entire universe was given by the time it took for a ray of light to circumnavigate it, which he estimated to 8,700 years. During its travel round the world the light would become dimmer because of absorption, and by taking into account the motion of the solar system he arrived at a loss in light intensity corresponding to 13 magnitudes.

---

[121]  Harzer 1908a, p. 248. See also the separately published Harzer 1908b.
[122]  Schwarzschild 1900, p. 346.
[123]  Seeliger based his model of the Milky Way universe on elaborate statistical analyses of star counts and stellar magnitudes. His model was similar to what in the history of astronomy is known as the "Kapteyn universe," a reference to the Dutch astronomer Jacobus Kapteyn. For details on the Seeliger-Kapteyn universe, see Paul 1993.

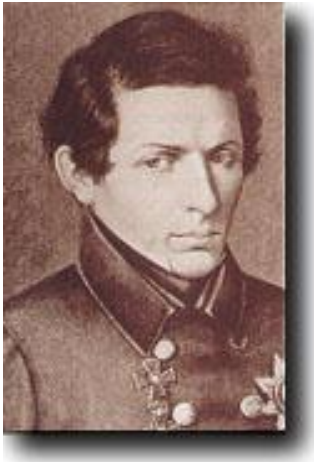 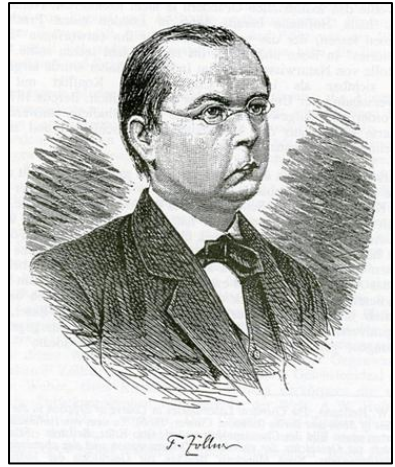

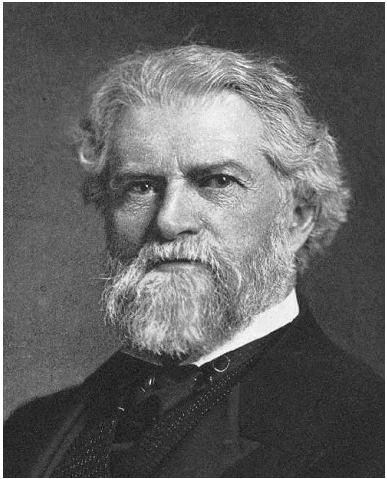 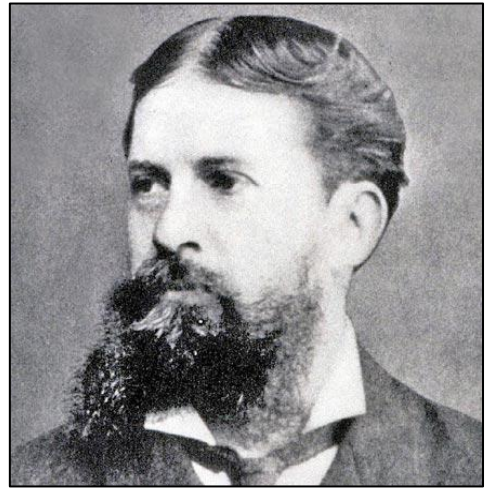

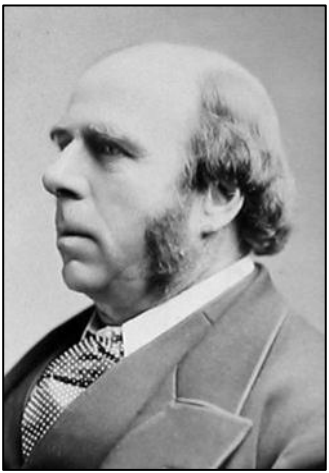 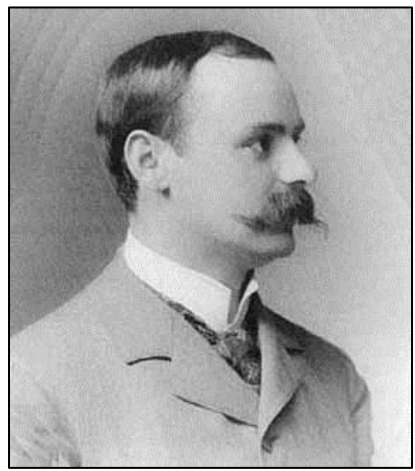

Pioneers of non-Euclidean astronomy: N. I. Lobachevsky (1792-1856); J. K. F. Zöllner (1834-1882); S. Newcomb (1835-1909); C. S. Peirce (1839-1914); R. S. Ball (1840-1913); K. Schwarzschild (1873-1916)

This was a more realistic value than Schwarschild's 40 magnitudes, yet it was sufficient to make the counter-sun invisible to our telescopes.

Harzer took the model of a closed stellar universe no less seriously than Schwarzschild, but of course he realized that it was hypothetical and lacked the support of solid observational evidence. Consequently, his conclusion was cautious: "This picture includes no features that can be characterized as improbable. … But the picture speaks of the *possibility* of the finite space only, not of its *reality*, and as yet we have no evidence for this reality."[124] Things had not really changed since the days of Zöllner.

Among the few mathematicians who expressed an interest in the consequences of non-Euclidean geometry in the realm of astronomy was the Briton William Barrett Frankland, who in 1910 published a small book in which he examined problems of planetary motion within this framework. For an elliptic space he obtained results that only differed inconceivably from those of the Newtonian theory formulated in ordinary Euclidean space.[125] Three years later, in an address to the London branch of the Mathematical Association, he pointed out that Riemannian geometry implied a space of finite extent, and that such a space would have some remarkable features. As earlier noted by Newcomb, Schwarzschild and others, a star should be visible in two opposite directions, and, disregarding absorption in interstellar space, the antipodal image should even be as bright as the real one. By taking into account the finite velocity of light, Barrett arrived at the more satisfactory result that the antipodal image of the sun would be much dimmer, even if space were perfectly transparent.[126]

---

[124] Harzer 1908a, p. 266.
[125] Frankland 1910, pp. 62-69.
[126] Frankland 1913.

## 8. Towards Einstein

During the early part of the twentieth century astronomers were well aware of the possibility of space being non-Euclidean, but it was considered a remote possibility only and nothing to keep them awake at night. To mention just one example, the 1911 edition of the recognized *Newcomb-Engelsmann Populäre Astronomie* included an account of Riemann's closed space. "Although this interpretation of the finite space transcends our conceptions, still it does not contradict them," the readers were informed. "However, what experience can tell us about it is limited by the fact that the entire observable universe may only be a small fraction of the entire finite space."[127] Characteristically, the account was placed in a footnote. Most books on cosmology did not mention the possibility of curved space, but presented the world as consisting of a huge conglomerate of stars, essentially the Milky Way. They were silent about space in any other capacity than an inert container of stars and other celestial bodies.

The generally accepted picture of the universe about 1915, the year that Einstein introduced his general theory of relativity, was a huge stellar system of an ellipsoidal form, with the density of stars diminishing with increasing distance from the centre. The dimensions were of the orders 50,000 light years in the galactic plane and 5,000 light years towards the galactic poles. The material universe was usually considered a finite stellar system in the infinite Euclidean space, and what might be beyond the stellar system was left to speculation. It might be empty space or some ethereal medium, in any case it was regarded as irrelevant from an astronomical point of view. The suggestion of Schwarzschild and Harzer of a closed space filled with stars had the advantage that it did away with the infinite empty space, but it made almost no impact on mainstream astronomy. The cosmological problem that moved to the forefront of astronomy in the 1910s was concerned with the size of the Milky Way system and the question of whether the spiral nebulae were external objects or belonged to the Milky Way. This was a problem in which the geometry of space was considered irrelevant. It was eventually solved in the mid-1920s, and then by purely observational means.

---

[127] Kempf 1911, p. 664. The widely acclaimed *Newcomb-Engelsmann Populäre Astronomie*, first published in 1905, was a much enlarged German translation of Newcomb's *Popular Astronomy* edited and updated by leading German astronomers.

While a few ideas about the finite universe were based on the notion of curved space, most were not. Thus, based on the paradoxes of Olbers and Seeliger, in 1896 the prominent Swedish astronomer and cosmologist Carl Charlier argued that the universe must be finite in space. Since he assumed the world to have existed in an eternity, he was then faced with the problem that the heat death has not already occurred. To escape the dilemma, he suggested that the stars are not uniformly distributed, and with this saving operation he concluded that "the world is finite in space."[128] However, what he had in mind was not closed space, but that the stellar system only filled a finite part of the infinite space. When Charlier developed his idea into a theory of an infinite, fractal and hierarchic universe, which he did in 1908, he continued to think of space as star-filled space. He defended his model as late as 1925, now arguing against a finite extension of the universe and without considering Einstein's cosmological model. He did refer to it, but only indirectly and only to reject it. "You know," he said, "that there also are speculative men in our time who put the question whether space itself is finite or not, whether space is Euclidean or curved (an elliptic or hyperbolic space)."[129] Although admitting that "such speculations lie within the domain of possibility," he found them to be unconvincing and just that – speculations.

---

[128]  Charlier 1896, p. 488.
[129]  Charlier 1925, p. 182. On Charlier's hierarchical universe and related ideas, see also Selety 1922.

After having studied Einstein's general theory of relativity, Schwarzschild quickly realized that the relativistic field equations have a solution corresponding to a closed universe with an elliptic geometry. By that time, in early 1916, Einstein had not yet begun investigating the cosmological consequences of his theory. Schwarzschild wrote him: "As concerns very large spaces, your theory has a quite similar position as Riemann's geometry, and you are certainly not unaware that one obtains an elliptic geometry from your theory if one puts the entire universe under uniform pressure."[130] When the Dutch astronomer Willem de Sitter more than a year later pointed out the distinction between spherical and elliptic space to Einstein, he referred to Schwarzschild's 1900 paper, which thus came to play a role in the origin of relativistic cosmology. As de Sitter noted, in agreement with Schwarzschild, although the spherical space is easier to visualize, "The elliptical space is, however, really the simpler one, and it is preferable to adopt this for the physical world." His careful discussion of "the spherical space, or space of Riemann, and the elliptical space, which has been investigated by Newcomb" suggests that the difference between the two types of closed space was not generally appreciated by physicists and astronomers at the time.[131]

The influence of Schwarzschild's old paper is also visible in de Sitter's estimate of the curvature of his empty and hyperbolic relativistic universe, known as the B model (contrary to Einstein's A model). Concerning the limit of stellar parallaxes, $p_{min} \cong 0''.005$, he said that "The limit found by Schwarszschild still corresponds to our present knowledge." He further referred to Schwarzschild's discussion of the absorption of light in elliptic space in order to avoid the problem of "an image of the back of the sun." Using recent data for the interstellar absorption, he estimated a value of $R > 2.5 \times 10^{11}$ AU for the Einstein universe.[132]

---

[130]  Schwarzschild to Einstein, 6 February 1916, in Einstein 1998a, document 188.
[131]  De Sitter 1917b, pp. 7-8, and similarly in de Sitter 1917a, pp. 231-234. See also Schemmel 2005, pp. 470-471, and de Sitter to Einstein, 20 June 1917, in Einstein 1998b, document 355.

Einstein's considerations eventually led him to the cosmological field equations including the cosmological constant $\Lambda$ that he thought was needed to keep the closed universe in a stationary state. His model of the universe was four-dimensional in space-time, and, satisfying the requirements of homogeneity and isotropy, with its metric being separable in the three space coordinates and the one time coordinate. It was "a self-contained continuum of finite spatial (three-dimensional) volume," as he wrote in his seminal paper of 1917.[133] Although the curvature of space would vary locally in time and space in accordance with the distribution of matter, he considered spherical space to be a good approximation on a cosmological scale. Einstein's theory resulted in definite formulae relating the mass and volume of the universe to the radius of curvature $R$, such as

$$M = 2\pi^2 \rho R^3 \, ,$$

However, given the uncertainty of the average density of matter $\rho$ this was of little help. In correspondence with his friend Michel Besso of March 1917, he suggested that $R \approx 10^7$ light years, a value which was based on the much too high estimate $\rho \approx 10^{-22}$ g cm$^{-3}$. Some months later he repeated the suggestion in a letter to de Sitter, but he wisely decided not to publish it.[134]

---

[132] De Sitter 1917b, p. 25. See also Peruzzi and Realdi 2011, pp. 667-668.
[133] Einstein 1923, p. 180.
[134] Einstein-Besso correspondence, March 1917, and Einstein to de Sitter, 12 March 1917, in Einstein 1998a, pp. 400-413.

Let me end by briefly recalling Einstein's brilliant address on geometry and experience that he gave to the Prussian Academy of Sciences in early 1921. On this occasion he distinguished between what he called "practical geometry" and "purely axiomatic geometry," arguing that while the first version was a natural science, the second was not. "The question whether the universe is spatially finite or not seems to me an entirely meaningful question in the sense of practical geometry," he said. "I do not even consider it impossible that the question will be answered before long by astronomy." Indeed, without this view of geometry, he continued, "I should have been unable to formulate the theory of [general] relativity."[135] Einstein elaborated on his view concerning physics and geometry in an article four years later, where he repeated that, in itself, geometry does not correspond to anything experienced. It only does so if combined with the laws of mechanics and optics, or with other laws of physics.[136]

Incidentally, Einstein's suggestion of 1921 that astronomical observations would soon reveal whether or not cosmic space is curved, turned out to be unfounded. Still in 1931, after he had accepted the expansion of the universe, he stuck to a closed universe, but the following year he changed his mind. In the important model he proposed jointly with de Sitter, he admitted that "There is no direct observational evidence for the curvature, … [and] from the direct data of observation we can derive neither the sign nor the value of the curvature." For reasons of simplicity, the Einstein-de Sitter model therefore described the universe "without introducing a curvature at all."[137] As far as the curvature of space was concerned, the Einsteinian revolution in cosmology did not change much.

---

[135] Einstein 1982, p. 239 and p. 235. It was in this address that Einstein stated his famous dictum that, "as far as the propositions of mathematics refer to reality, they are not certain; and as far as they are certain, they do not refer to reality." For an in-depth analysis of Einstein's lecture, see Friedmann 2002.

[136] Einstein 1925.

## 8. Conclusion

Whereas non-Euclidean geometry flourished as a mathematical research field in the last half of the nineteenth century (see the figure on p. 8), its connection to the real space inhabited by physical objects was much less cultivated. The large majority of mathematicians did not care whether real space was Euclidean or not; and those who did care only dealt with the subject in a general and often casual way, avoiding to deal seriously with the possibility of determining a space curvature different from zero. After all, that was supposed to be the business of the astronomers. While some mathematicians, following Poincaré, declared the problem meaningless, others admitted that in principle space might be curved – but in principle only – and left it by that.

Astronomers had their own reasons to ignore non-Euclidean geometry, and especially as applied to space on the cosmological scale. Lack of awareness of the new forms of geometry, or lacking mathematical competence, was not generally among the reasons: many astronomers had a strong background in mathematics and were conversant with the technicalities of non-Euclidean geometry. But while the motion and properties of celestial bodies were definitely the business of the astronomers, the space in which the bodies move was not seen as belonging to the domain of astronomy. It was a kind of nothingness that philosophers could speak of, and did speak of. The recognized mainstream astronomer Newcomb probably spoke for the majority of his colleagues when he warned against "the tendency among both geometers and psychologists to talk of space as an entity in itself."[138] To arouse interest in the astronomical community, theories of non-Euclidean space would have to be observationally testable or offer opportunities for solving problems of astronomical relevance. They scored badly on both accounts.

---

[137] Einstein and de Sitter 1932.

In a few cases problems of astronomy, such as the anomalous motion of Mercury, were approached by means of techniques based on the hypothesis of curved space, but this soon turned out to be a dead end. On the other hand, even though non-Euclidean geometry might have little or no explanatory force, there was the possibility that it could be verified by measurements. While it could never be proved that space is Euclidean, it could conceivably be proved that it is not. A few astronomers and other scientists did take an interest in this line of reasoning going back to Lobachevsky. While many standard histories jump from Lobachevsky to Schwarzschild, we have seen that the interest was more widespread, including scientists such as Ball, Newman, Peirce and Barbarin. However, while in the early years of the twentieth century it was realized that the curvature of space was indeed measurable, it was also realized that the kind of upper bound for the curvature that measurements allowed was ineffective to distinguish curved from flat space. Given this situation, no wonder that astronomers saw no reason to abandon the intuitively pleasing Euclidean space that had served their science so well in the past. Even should space be curved, the curvature radius would be so large that for all practical purposes it was infinite, that is, space could be considered Euclidean. So why bother?

Another problem of astronomical or rather cosmological relevance that might have induced astronomers to consider the curved nature of space was related to the question of whether space is finite or infinite in extent. The question might be seen as merely philosophical, and it often was, but it had observational consequences related to Olbers' optical paradox and Seeliger's gravitation paradox. Yet only in one case, Zöllner's discussion of 1872, was the problem of the shining night sky approached by arguing that the stellar universe might be closed in accordance with Riemann's hypothesis. The next time the closed universe turned up as more than a commentary made in passing, was in Schwarzschild's article of 1900, nearly thirty years later.

---

[138] Newcomb 1898, p. 5.

In conclusion, it seems that the main reason for the astronomers' reluctance to consider the consequences of space being non-Euclidean was just this: they had no need for the hypothesis. An additional reason may have been the popular association between non-Euclidean geometries and four-dimensional hyperspace. In spite of its lack of justification, the fourth spatial dimension remained alive and well, and not only in spiritualist circles. "With the exception of Zöllner," Russell remarked, "I know of no one who has regarded the fourth dimension as required to explain phenomena."[139] He might have said the same about the non-Euclidean spaces of three dimensions.

---

[139] Russell 1897, p. 53. Some scientists, apart from Zöllner, thought that the fourth dimension might be of use in explaining phenomena. Newcomb was open for the idea, and so were a few other scientists and writers (Beichler 1988). One of them was Charles Hinton, and another was T. Proctor Hall, an American mathematician, who discussed the possibility in a paper in *Science*. It probably did not help his cause that of the phenomena that might be explained by means of the fourth dimension, he singled out the appearance of ghosts (Hall 1892).